\documentclass[
reprint,
superscriptaddress,
nofootinbib,
amsmath,amssymb,
aps,
longbibliography,
prx,
]{revtex4-2}

\usepackage{graphicx}
\usepackage{mathtools}
\usepackage{tikz}

\usepackage{hyperref}
\hypersetup{
breaklinks=true,
colorlinks=true,
citecolor=blue,
linkcolor=black,
filecolor=black,
urlcolor=blue
}
\IfFileExists{lmodern.sty}{\usepackage{lmodern}}{}

\def\be{\begin{equation}}
\def\ee{\end{equation}}

\DeclareMathOperator{\Tr}{Tr}
\DeclareMathOperator{\Id}{Id}
\DeclareMathOperator{\sech}{sech}

\newcommand{\llangle}{\langle\!\langle}
\newcommand{\rrangle}{\rangle\!\rangle}

\begin{document}

\title{Exactly solvable non-unitary conformal interfaces in unitary CFTs}

\author{Qicheng Tang}
\email{qctang@gatech.edu}
\affiliation{School of Physics, Georgia Institute of Technology, Atlanta, GA 30332, USA}

\author{Zixia Wei}
\affiliation{Society of Fellows, Harvard University, Cambridge, MA 02138, USA}

\author{Xueda Wen}
\affiliation{School of Physics, Georgia Institute of Technology, Atlanta, GA 30332, USA}

\begin{abstract}

We construct directly on the lattice a class of non-unitary interfaces that are both exactly conformal and exactly solvable, and establish their corresponding boundary and interface conformal field theory (CFT) descriptions. The construction is obtained by analytically continuing the scattering data of known exact unitary conformal interfaces on the lattice, yielding an $SL(2,\mathbb C)$-parametrized family, which is non-compact and breaks probability-current conservation. 
Exploiting the exact lattice-continuum correspondence, we derive the conformal boundary states in the folded picture. We show that a proper definition of the Hilbert space in the closed-string channel requires the incoming and outgoing boundary states to be specified independently by boundary data associated with a pair of dual biorthogonal bases, in close analogy with the right and left eigenvectors of a non-Hermitian Hamiltonian. This requirement determines a consistent CFT construction of non-unitary boundaries and interfaces, and leads to a non-unitary generalization of the conventional Cardy's condition for unitary boundary CFT. 
Beyond their formal construction, these non-unitary interfaces are shown to exhibit logarithmic entanglement scaling governed by an effective central charge that is generally complex. For the $SU(1,1)$ subclass, the effective central charge remains real but grows without bound as the transmission coefficient increases. This result is demonstrated through analytical and numerical lattice calculations, as well as an interface CFT analysis in the unfolded picture. 
Finally, we present a general CFT analysis of a class of global quantum quenches whose initial states are prepared with non-unitary boundaries. We relate their effective temperature to the conformal dimension of the boundary-condition-changing operators associated with non-unitary boundary conditions.

\end{abstract}

\maketitle

\tableofcontents

\section{Introduction}

Impurities and interfaces are ubiquitous in physical systems. They give rise to rich phenomena across a wide range of energy scales and disciplines, from unconventional transport across impurities and junctions in condensed matter physics \cite{kondo1964, andreev1964thermal, Leggett_1981_dissipation_tunnel, Kane_Fisher_1992_transport_luttinger} to brane-induced modifications of geometry and vacuum structure in string theory \cite{Polchinski_textbook}. A particularly important class of examples arises when the bulk systems on the two sides of an interface are critical. In this setting, the infrared bulk physics is governed by conformal field theories (CFTs)~\cite{cft_yellow_book}, whereas the degrees of freedom near the interface may undergo their own renormalization-group (RG) flow and reach an interface fixed point described by a \emph{conformal interface}~\cite{Affleck1994folding,1996_Oshikawa,1997_Oshikawa,
2001_Bachas_Boer_Dijkgraaf_Ooguri, Frohlich2004_KW_duality, Quella_2006_transmission_defect, Brunner_2008_RG_DW, 2008_Sakai, Eisler_2010_EE_fermion_defect,Eisler_2012_defect_boson, Brehm2015_EE_interfece_ising, 
2018_Karch,Bachas2020,2020_Meineri,Rogerson2022,2023_Karch,2023_Tang_Wei,2024_Karch}. Such interfaces are of both conceptual and practical importance: they provide diagnostic probes of universal bulk conformal data and of gapless edge modes in topological phases~\cite{Zou_2020_markov_gap, Siva_2021_tripartite_edge, Liu_2021_multipartite_vertex, yuya2022_reflected, 2023_Tang_Wei, Liu_2023_multipartite}, while also supporting distinct critical phenomena of their own~\cite{Affleck1995_kondo,1996_Oshikawa,1997_Oshikawa,Kane_Fisher_1992_transport_luttinger}. They can also appear as RG interfaces connecting distinct fixed points or encoding flows between different bulk theories, thereby offering a useful way to organize the theory space of quantum critical points~\cite{Brunner_2008_RG_DW, Gaiotto2012_RG_DW, Konechny_2014_RG_defect, Brunner_2016_transmission, Cardy2017_boundary_bulkRG, Konechny2021_RGinterface}.

In two dimensions, a conformal interface is an object that satisfies a local gluing condition: the appropriate component of the stress-energy tensor must be continuous across it, which ensures the preservation of conformal transformations that are compatible with the presence of the interface. However, this conformal gluing condition is only a local symmetry constraint and does not by itself specify the full interface data, including the spectrum of interface operators, bulk-to-interface OPE coefficients, or the scattering data governing reflection and transmission across the interface. If the bulk theories possess additional structures, such as global symmetries or chiral algebras, one may further impose gluing conditions requiring the interface to preserve a chosen common symmetry or sub-algebra. Such requirements define more constrained classes of interfaces rather than following from unitarity or conformal invariance alone. Moreover, a consistent interface must also satisfy global consistency constraints~\cite{Cardy_1989, Lewellen_1992_sewing_boundary}, most notably open-closed duality (also known as the Cardy's condition~\cite{Cardy_1989}), which relates the closed-channel description of a cylinder amplitude involving the interface to its open-channel interpretation in terms of interface degrees of freedom as well as the underlying Virasoro algebra of the bulk theory. The resulting classification problem is subtle even when the bulk theories are rational~\cite{Quella_2006_transmission_defect, Brunner_2008_RG_DW}. Although the folding trick maps an interface to a boundary condition in a product theory~\cite{Affleck1994folding,1997_Oshikawa,2001_Bachas_Boer_Dijkgraaf_Ooguri}, existing classifications of boundary conditions in rational conformal field theory do not directly yield a classification of interfaces, since the relevant boundary problem in the folded description need not be rational. Thus, the general classification of conformal interfaces and the identification of their microscopic realizations remain highly nontrivial.

Non-unitarity sharpens these difficulties. It has long appeared in critical phenomena, ranging from classic examples such as the Yang-Lee edge singularity, percolation, and logarithmic CFTs \cite{MichaelFisher_imaginary_phi3,Essam_1980_percolation,Cardy_1991_percolation,1993_Gurarie,2003_Flohr,2013_logCFT_review} to more recent developments involving non-Hermitian criticality~\cite{
2019_Chang,2026_Hsieh,2025_Barad,2026_Karch_Wang,2026_Tang}, complex CFTs~\cite{Rychkov_2018_weak_1,2018_Walking2,Ma2018,Faedo_2019,Benini_2019,2024_Zhu,Jacobsen2024_complex_CFT,2023_Haldar,2025_Zhu,2025_Kohei,Kohei_2026_complex_nonlinear_sigma,2026_Kohei,2026_Yang, Jacobsen2026_complex_CFT}, monitored quantum dynamics~\cite{Chen_2020_nonunitary_random_fermion, Tang_2021_nonunitary_dynamics}, and measurement-induced criticality~\cite{Skinner2018_MIPT, Li2018_MIPT, Li_2020_bcc_mesurement}. These examples show that non-unitary critical behavior is not merely a formal extension of unitary critical phenomena, but can arise in physically motivated systems or in effective descriptions of equilibrium and non-equilibrium criticality. At the same time, the theory of non-unitary boundaries and interfaces remains considerably less developed~\cite{
Gaberdiel2001_SL2C_boundary, Gaberdiel2002_Dbrane_SL2C,
Hasselfield2005_sineGordon_SL2C_boundary,2008_Korff,2025_Zhu,2025_Li,2025_Zhou,2025_Ryu}. Boundary and interface CFTs associated with non-unitary bulk theories arise naturally, and some special cases are under good algebraic control~\cite{Quella_2006_transmission_defect}. However, when the bulk theory is already non-unitary, it is difficult to disentangle the constraints intrinsic to the boundary or interface from the non-unitary features already present in the bulk. Moreover, many structural arguments familiar from unitary boundary and interface CFTs, including those based on Hermiticity, positivity, Hilbert-space interpretation, and boundary or interface RG flows, require substantial reformulation.

These considerations motivate a complementary route in which the two-dimensional bulk remains at a conventional unitary critical point, while the departure from unitarity enters through boundary or interface couplings rather than through the bulk theory itself. The purpose of this restriction is not merely to simplify the problem or to generate formal examples, but to place the interface in a setting where the bulk properties are controlled by a comparatively well-understood unitary theory. In this way, one can ask more sharply what additional structures or constraints are needed for non-unitary boundary or interface data to define a genuine conformal fixed point. This question is physically natural in situations where only a local region is coupled to noise, dissipative channels, or measurement devices. After coarse-graining, conditioning on measurement outcomes, or integrating out auxiliary degrees of freedom, such setups may be effectively described by non-unitary interface terms. These mechanisms motivate the possibility of departures from unitarity introduced at the level of local microscopic couplings near the interface, without requiring the bulk critical theory away from it to be non-unitary. The resulting theoretical problem is therefore twofold: to understand whether local critical systems with such boundary/interface couplings can realize non-unitary conformal boundary or interface fixed points, and to use their existence and internal structure to infer what consistency conditions a broader theory of non-unitary boundary/interface CFT should satisfy. Non-unitary boundary conditions in unitary CFTs also appear in the AdS/CFT context realized by AdS$_3$ gravity coupled to AdS$_2$ brane with imaginary-valued scalar fields \cite{2026_Takayanagi}  and dS$_2$ branes \cite{HJW26}.

This perspective also clarifies why the problem is subtle. Relaxing unitarity at the interface weakens the usual link between algebraic consistency and physical realizability. In unitary boundary and interface CFTs, open-closed duality is not merely a formal identity of cylinder amplitudes: it is tied to a Hilbert-space interpretation in which spectra, operator algebras, inner products, and RG flows from microscopic models are constrained by Hermiticity, positivity, and locality. Once unitarity is relaxed at the interface, this structure is no longer canonical. For example, a non-Hermitian open-channel Hamiltonian may be naturally described in terms of distinct left and right eigenstates with a biorthogonal pairing, rather than a single orthonormal basis, and it is not generally clear what closed-channel state space, pairing, or boundary/interface state should encode this data. Thus, even when open-closed duality can be imposed as an algebraic constraint on cylinder amplitudes, its physical interpretation and its ability to distinguish microscopically realizable interface fixed points from formal candidates remain open questions.

From the continuum field-theoretic point of view, this absence of a canonical state-space interpretation makes non-unitary interfaces intrinsically ambiguous unless an additional physical selection principle is supplied. One can formally enlarge the space of candidate interface data by analytically continuing parameters of known solutions, complexifying boundary or interface couplings, or allowing non-unitary representations in the boundary/interface gluing conditions. Such procedures may produce candidate scale-invariant boundary states, interface spectra, or reflection and transmission data. However, by themselves they do not determine whether the resulting objects correspond to genuine interface fixed points of local critical systems or to artifacts of an enlarged algebraic parameter space. Microscopic realizations in local systems with unitary bulk critical behavior therefore play a particularly important role: they provide a way to distinguish realizable non-unitary interfaces from formal continuum candidates, and at the same time help reveal the consistency conditions that a broader theory of non-unitary boundary and interface CFTs should satisfy.

\section{Overview of main results}

In this work, we address this problem by constructing a class of lattice models with localized non-Hermitian defects that realize exact non-unitary conformal interfaces in otherwise unitary critical systems. We develop the construction in two standard free critical settings: a free-fermion chain whose scaling limit is the free Dirac CFT, and a harmonic-oscillator chain whose scaling limit is the free scalar CFT. In the main text, we focus on the Dirac fermion case, since it gives the most economical route to the scattering, entanglement, boundary-state, interface-operator, and biorthogonal open-closed structures that are central to our analysis for building up the proper boundary/interface CFT description for non-unitary conformal interfaces. The bosonic realization is worked out separately in the Appendix~\ref{Appendix:Boson} because it follows the same organizing principle and shares many technical steps with the fermionic case, not because it is merely auxiliary. Rather, it provides a parallel microscopic realization of exact non-unitary conformal interface of the same mechanism.

The key feature of these models is that the single-particle scattering matrix at the defect is exactly mode independent throughout the lattice spectrum, rather than becoming scale invariant only asymptotically in the low-energy limit. In this sense, the scale-invariant scattering characteristic of a conformal interface is implemented directly and exactly at the microscopic level. The lattice construction therefore does more than provide a solvable regularization of a continuum fixed point: it gives concrete microscopic realizations of non-unitary conformal interfaces embedded in otherwise unitary critical systems, and avoids relying solely on potentially ambiguous non-unitary RG flows.

In the Dirac fermion case, the known unitary conformal interface can be generated from the defect-free Hamiltonian by a unitary transformation controlled by the interface parameter. In Sec.\ref{Sec:Lattice}, we show that, after analytically continuing this parameter to complex values, the same construction becomes a generally non-unitary similarity transformation. Away from singular points, this transformation is isospectral to the Hermitian defect-free Hamiltonian while keeping the single-particle scattering matrix mode independent. The non-Hermitian nature of the defect, however, makes the scattering problem intrinsically biorthogonal: right and left eigenmodes are associated with distinct scattering matrices, and both are needed to characterize the interface. In the continuum free Dirac theory, this construction extends the familiar $SU(2)$ family of unitary conformal interfaces to an $SL(2,\mathbb{C})$ family, in which probability-current conservation is no longer imposed.

Guided by this exact microscopic construction, in Sec.\ref{Sec:CFT}, we develop the corresponding continuum description in both the folded and unfolded pictures, including explicit boundary states and interface operators. The lattice realization plays an important role here because it selects how the non-unitary continuum data should be assembled. In particular, boundary data built only from right eigenmodes or only from left eigenmodes do not lead to a closed-channel cylinder amplitude with the correct open-channel interpretation. Open-closed duality instead requires the incoming and outgoing boundary/interface states to be constructed from biorthogonal pairs of scattering data. This gives a concrete formulation of Cardy's condition in the present non-unitary setting, and clarifies the role of the closed-channel state space and pairing when the open-channel Hamiltonian is non-Hermitian. 

The same scattering data also determine correlation functions and entanglement across the interface, for which we discussed in Sec.\ref{Sec:Lattice} and Sec.\ref{Sec:CFT}. For the biorthogonal ground state, the entanglement entropy exhibits the expected logarithmic scaling. Its logarithmic coefficient can be expressed in terms of an effective central charge, whose analytic form is identical to that of the unitary conformal interface, with the interface parameter analytically continued to the non-unitary regime. This effective central charge is therefore generally complex, reflecting the non-unitary nature of the interface fixed point. In a distinguished $SU(1,1)$ subfamily inside $SL(2,\mathbb{C})$, the same closed-form expression remains real but becomes unbounded as the interface parameter is increased, a behavior associated with the amplification or attenuation of single-particle wavefunctions across the interface. The result is obtained consistently from the exact lattice calculation, numerical simulations, and the continuum replica computation using the non-unitary interface operator in the unfolded interface-CFT description.

Finally, the microscopic realizability of non-unitary conformal boundary/interface data gives a firmer basis for using non-unitary conformal boundary states in broader CFT discussions. As an illustration, in Sec.\ref{Sec:Quench}, we revisit quantum quenches to criticality at the level of general boundary CFT, without relying on a specific microscopic quench realization. In the standard Calabrese-Cardy framework, an initial state with a finite correlation length is represented by a conformally cooled boundary state, and the cooling length sets the effective temperature of the quench dynamics. For a non-unitary boundary state, however, the relevant Euclidean amplitude involves both the boundary state and its conjugate or dual, and the boundary-condition-changing (bcc) operator between them can contribute to the effective temperature of the quench. Since the scaling dimension of this operator may be negative in a non-unitary theory, this contribution can increase the effective temperature, or equivalently decrease the effective inverse temperature, therefore serves as a candidate for explaining the disagreement between CFT prediction and general observations in quantum many-body systems. Moreover, the same agreement, i.e. by scanning the physical effect of the non-unitary bcc operators, is potentially useful for inferring non-unitary boundary data in more general critical systems.

\section{Insights from lattice model}
\label{Sec:Lattice}

\subsection{Free Dirac fermion with unitary conformal interface on an interval}

Our starting point is a free fermionic chain of length $2L$ with a conformal interface located on the central bond. The Hamiltonian is
\begin{equation}\label{eq:defect_Ham}
\hat{\widetilde{\mathbf H}} = \sum_{m,n = 1}^{2L} \widetilde{H}_{m,n} c_m^\dagger c_n ,
\end{equation}
The non-zero elements of the single-particle kernel $\widetilde{H}$ are
\begin{equation}
\begin{aligned}\label{eq:matrix_element_defect_Ham_open}
\widetilde{H}_{m, m+1} = \widetilde{H}_{m+1, m} & = \begin{cases}
-{1}/{2} , & m \neq L, \\ -{\lambda}/{2} , & m = L,
\end{cases}
\end{aligned}
\end{equation}
and
\be
- \widetilde{H}_{L, L} = \widetilde{H}_{L+1, L+1}  = {\mu}/{2} = \sqrt{1 - \lambda^2} / 2,
\ee
where $\lambda$ is a parameter that characterizes the interface, $\mu^2+\lambda^2=1$ is the condition that makes the interface exactly conformal on lattice, i.e., the scattering data of each single-particle mode is independent of the mode index,
which is related to the energy of the scattering modes.

This lattice Hamiltonian has been extensively studied in the Hermitian regime $\lambda \in [0, 1]$ \cite{Eisler_2010_EE_fermion_defect, Eisler_2012_defect_boson,Eisler_2012_evo_defect}. In this case, the eigenvalue decomposition of $\widetilde{H}$ is
\begin{equation}\label{eq:defect_unitary_eigen}
\widetilde{H} \widetilde{\phi}_k = E_k\widetilde{\phi}_k,
\end{equation}
where the eigenvalues are identical to the standard defect-free open chain
\begin{equation}\label{eq:defect_unitary_eigenvalue}
E_k = - \cos \theta_k \,, \quad \theta_k = \frac{k \pi}{2L+1} \,, \quad k=1,\cdots,2L.
\end{equation}
The corresponding eigenfunctions, however, are modified by the presence of the interface and take the form
\begin{equation}\label{eq:defect_wf}
	\widetilde{\phi}_k(m) = \begin{cases}
		\alpha_k\, \phi^0_k(m),
		& \quad 1 \leq m \leq L, \\
		\beta_k\, \phi^0_k(m),
		& \quad L+1 \leq m \leq 2L. \\
	\end{cases}
\end{equation}
Here the coefficients satisfies
\begin{equation}\label{eq:prefactor_defect_wf}
	\alpha_k^2 = 1 - (-1)^k \mu , \quad \beta_k^2 = 1 + (-1)^k \mu,
\end{equation}
and $\phi^0_k(m)$ is the normalized eigenfunction for a standard defect-free open chain
\begin{equation}\label{eq:defect_free_wf_open}
\phi^0_k(m) = \sqrt{\frac{2}{2L+1}} \sin \left( m \theta_k \right).
\end{equation}

To demonstrate the inserted interface is exactly conformal, one needs to evaluate the single-particle scattering matrix $S(k)$ for each mode labeled by the wavenumber $k$. Expanding the eigenvalue decomposition of $\widetilde{H}$ in \eqref{eq:defect_unitary_eigen} yields the following interface equations
\begin{equation}\label{eq:interface_eq_eigen}
\begin{aligned}
\phi_k(L-1) + \mu \phi_k(L) + \lambda \phi_k(L+1) &= -2E_k\phi_k(L) ,
\\
\lambda \phi_k(L) - \mu \phi_k(L+1) + \phi_k(L+2) &= -2E_k\phi_k(L+1) .
\end{aligned}
\end{equation}
For later convenience, we denote the bulk theories on the left and right sides of the interface by
CFT$^{\rm I}$ and CFT$^{\rm II}$, respectively.
The incoming and outgoing modes at the interface can be defined in terms of the single-particle wavefunction as
\begin{equation}
\label{eq:relation_modes_wavefunction}
	\begin{aligned}
		a_k^{\rm I, in} = a_k^{{\rm I}, R} &= e^{i \frac{3}{2} \theta_k} \left[ \phi_k(L) - e^{-i \theta_k} \phi_k(L-1) \right] \\
		a_k^{\rm I, out} = a_k^{{\rm I}, L} &=  -e^{-i \frac{3}{2} \theta_k} \left[ \phi_k(L) - e^{i \theta_k} \phi_k(L-1) \right] \\
		a_k^{\rm II, in} = a_k^{{\rm II}, L} &= e^{i \frac{3}{2} \theta_k} \left[ \phi_k(L+1) - e^{-i \theta_k} \phi_k(L+2) \right] \\
		a_k^{\rm II, out} = a_k^{{\rm II}, R} &= -e^{-i \frac{3}{2} \theta_k} \left[ \phi_k(L+1) - e^{i \theta_k} \phi_k(L+2) \right]
	\end{aligned}
\end{equation}
where superscripts $L$ and $R$
denote left- and right-moving modes, respectively.
By substituting the above relations to the interface equations in \eqref{eq:interface_eq_eigen}, one obtains
\begin{equation}
S(k) \,a_{\rm in} = a_{\rm out} \,, \quad
a_{\rm in} = \begin{pmatrix} a_k^{\rm I, in} \\[0.5ex] a_k^{\rm II, in} \end{pmatrix}, \, \,\,
a_{\rm out} = \begin{pmatrix} a_k^{\rm I, out} \\[0.5ex] a_k^{\rm II, out} \end{pmatrix},
\end{equation}
with the \emph{mode-independent} scattering matrix
\begin{equation}
S(k) = S = \begin{pmatrix}
\mu & \lambda \\ \lambda & -\mu
\end{pmatrix}.
\end{equation}
Pictorially, the scattering matrix describes the following scattering process:
\be
\begin{tikzpicture}[x=0.75pt,y=0.75pt,yscale=-0.65,xscale=0.65]

\draw [color={rgb, 255:red, 74; green, 74; blue, 74 }  ,draw opacity=1 ][line width=1.5]    (260,41.35) -- (260.02,170.61) ;
\draw    (205.01,161.98) -- (260.01,105.98) ;
\draw [shift={(236.01,130.42)}, rotate = 134.48] [fill={rgb, 255:red, 0; green, 0; blue, 0 }  ][line width=0.08]  [draw opacity=0] (10.72,-5.15) -- (0,0) -- (10.72,5.15) -- (7.12,0) -- cycle    ;
\draw    (260.01,105.98) -- (315.01,49.98) ;
\draw [shift={(291.01,74.42)}, rotate = 134.48] [fill={rgb, 255:red, 0; green, 0; blue, 0 }  ][line width=0.08]  [draw opacity=0] (10.72,-5.15) -- (0,0) -- (10.72,5.15) -- (7.12,0) -- cycle    ;
\draw    (316,160.64) -- (260.01,105.98) ;
\draw [shift={(284.43,129.82)}, rotate = 44.31] [fill={rgb, 255:red, 0; green, 0; blue, 0 }  ][line width=0.08]  [draw opacity=0] (10.72,-5.15) -- (0,0) -- (10.72,5.15) -- (7.12,0) -- cycle    ;
\draw    (260.01,105.98) -- (204.02,51.32) ;
\draw [shift={(228.44,75.16)}, rotate = 44.31] [fill={rgb, 255:red, 0; green, 0; blue, 0 }  ][line width=0.08]  [draw opacity=0] (10.72,-5.15) -- (0,0) -- (10.72,5.15) -- (7.12,0) -- cycle    ;

\draw (200,140) node   {$a_k^{\rm I, in}$};
\draw (195,80) node   {$a_k^{\rm I, out}$};

\draw (330,140) node   {$a_k^{\rm II, in}$};
\draw (330,80) node   {$a_k^{\rm II, out}$};

\end{tikzpicture}
\ee

Moreover, a $U(1)$ topological defect can be fused onto this conformal defect without spoiling either exact solvability or conformality. For an open chain, this fusion amounts to introducing $U(1)$ phases in the interface hopping terms as follows,
\be
\label{H_tilde_U1}
\begin{aligned}
	\widetilde{H}_{L, L+1} & = -\lambda/2
	\quad \to \quad
	-e^{+i\Delta}\lambda/2, \\
	\widetilde{H}_{L+1, L} & = -\lambda/2
	\quad \to \quad
	-e^{-i\Delta}\lambda/2 ,
\end{aligned}
\ee
resulting in a new Hamiltonian kernel, denoted by $\widetilde{H}^{\,U(1)}$.
The eigenvalues of the defect Hamiltonian in \eqref{eq:defect_unitary_eigenvalue}, as well as the relation to the defect-free eigenfunctions in \eqref{eq:defect_wf}, retain the same form after this additional phase twist. The only modification is that the reference defect-free eigenfunctions are replaced by
\begin{equation}\label{eq:free_wf_U1}
	\phi_k^0(m) \to \phi_k^{0,U(1)}(m) =
	\begin{cases}
		\phi_k^0(m),
		& 1 \leq m \leq L, \\
		e^{-i\Delta} \phi_k^0(m),
		& L+1 \leq m \leq 2L.
	\end{cases}
\end{equation}
Here, $\phi_k^{0,U(1)}$ denotes the corresponding reference eigenfunction after fusing the additional $U(1)$ topological defect.
Consequently, the scattering matrix becomes
\be
S \to S_{U(1)} = \begin{pmatrix}
	\mu & e^{+i\Delta} \lambda \\
	e^{-i\Delta} \lambda & -\mu
\end{pmatrix} \in U(2),
\ee
thereby introducing a physically measurable transmission phase $\Delta$.
In the following, unless stated otherwise, $S$ denotes the $U(1)$-extended scattering matrix with $\det S=-1$, viewed as a subset of $U(2)$. Multiplication by an imaginary unit $i$ maps it to the standard $SU(2)$ representation~\footnote{In the continuum free-Dirac CFT, after folding the interface, a fermion-number-conserving boundary condition is described by a scattering matrix $S\in U(2)$. Quotienting out the overall $U(1)$ phase gives an $SU(2)$ matrix, characterized by the transmission amplitude, transmission phase, and reflection phase. In our exactly solvable lattice model, the reflection phase is fixed, so only the transmission amplitude $\lambda$ and the transmission phase $\Delta$ are tunable. }
\be
\begin{aligned}\label{eq:unitary_su2_scattering}
	\hat{S}_{\rm unitary} & = i S
	= \begin{pmatrix}
		i\mu & i e^{+i\Delta} \lambda \\ i e^{-i\Delta} \lambda & -i\mu
	\end{pmatrix}
	\\ & = \begin{pmatrix}
        i \cos\varphi & ie^{+i\Delta} \sin\varphi \\
		ie^{-i\Delta} \sin\varphi & -i \cos\varphi
	\end{pmatrix} \in SU(2),
\end{aligned}
\ee
where $\sin\varphi=\lambda \,,\, \cos\varphi=\mu \,,\,  \varphi \in [-\frac{\pi}{2},\frac{\pi}{2}]$.

The transmission and reflection coefficients of the conformal interface are given by
\begin{equation}
	\mathcal{T} = \lvert \lambda \rvert^2 = \lambda^2 , \quad
	\mathcal{R} = \lvert \pm \mu \rvert^2 = \mu^2,
\end{equation}
which are continuously tunable. They satisfy the unitarity condition
\begin{equation}
	\mathcal{T} + \mathcal{R} = \lambda^2 + \mu^2 = 1.
\end{equation}
In particular, unitarity implies the conservation of the probability current,
\begin{equation}\label{eq:unitary_current_perserve}
	a_{\rm out}^\dagger a_{\rm out} = {a}_{\rm in}^\dagger \,S^\dagger S \,{a}_{\rm in} = {a}_{\rm in}^\dagger {a}_{\rm in},
\end{equation}
which holds for any unitary interface satisfying $S^\dagger S=\Id$.

\subsection{Analytic continuation to non-unitary interface: from a similarity transformation}

We now analytically continue $\lambda$ from the Hermitian regime $\lambda \in [-1,1]$~\footnote{The region $\lambda\in[-1,0]$ can be obtained from $\lambda\in[0,1]$ by fusing a $U(1)$ topological defect with $\Delta=\pi$; in this case the Hamiltonian remains Hermitian.} to complex values, $\lambda \in \mathbb{C}$.

Before turning to the non-unitary interface, it is useful to recall the key observation underlying the exact solvability of the unitary interface: the defect Hamiltonian is related to the defect-free Hamiltonian by a similarity transformation,
\be
\widetilde{H}^{U(1)} = V H^{0,U(1)} V^{-1} \,,
\ee
where $\widetilde{H}^{U(1)}$ is defined through \eqref{H_tilde_U1}.
This equation relates the defected Hamiltonian to the defect-free one. In the Hermitian regime, allowing for a possible fused $U(1)$ topological defect, this matrix is obtained from the corresponding single-particle wavefunctions as
\begin{equation}
	V = \Phi_0^{U(1)} \left( \widetilde{\Phi}^{U(1)} \right)^{-1}
	= \Phi_0^{U(1)} \left[ \widetilde{\Phi}^{U(1)} \right]^\dagger .
\end{equation}
Here $\Phi_0^{U(1)}(m,k)=\phi_k^{0,U(1)}(m)$ and $\widetilde{\Phi}^{U(1)}(m,k)=\widetilde{\phi}{}_k^{\,0,U(1)}(m)$ are given in \eqref{eq:defect_wf} and \eqref{eq:free_wf_U1}. After some algebra, one finds that only the diagonal and anti-diagonal elements of $V$ are nonzero:
\begin{equation}
	\begin{aligned}\label{eq:element_orth_matrix_defect}
		V_{m, m} &= \frac{1}{2} \left( \sqrt{1 + \mu} + \sqrt{1 - \mu} \right) , \\
		V_{2L+1 - m, m} & = \frac{s(m) e^{-i\Delta\, s(m)}}{2} \left( \sqrt{1 + \mu} - \sqrt{1 - \mu} \right) ,
	\end{aligned}
\end{equation}
where $s(m)=1$ for $m=1,\cdots,L$ and $s(m)=-1$ for $m=L+1,\cdots,2L$. Its inverse also has only non-zero diagonal and anti-diagonal elements
\begin{equation}
		(V^{-1})_{m, m} = V_{m,m} \,, \quad
		(V^{-1})_{2L+1 - m, m} = - V_{2L+1 - m, m} \,.
\end{equation}

We now turn to the non-unitary conformal interface. The above expression for $V$ can be analytically continued to $\lambda\in\mathbb C$, yielding a Hamiltonian of the same local form as that in Eq.\eqref{eq:matrix_element_defect_Ham_open}.~\footnote{In the Hermitian regime $\lambda \in [-1,1]$, one has $V^{-1}=V^\dagger$. When $\Delta=0$, one further has $V^{-1}=V^T$; after analytically continuing to complex $\lambda$, $V$ becomes complex orthogonal.} Therefore, in the non-Hermitian regime $\lambda \in \mathbb{C} \setminus [-1,1]$, the spectrum remains real and identical to the defect-free case. Moreover, the non-Hermitian Hamiltonian is generally not parity-time symmetric, except on the real branches $\lambda\in(-\infty,-1)\cup(1,+\infty)$.

The biorthogonal right and left eigenfunctions of the non-hermitian defected Hamiltonian for $\lambda \in \mathbb C \setminus [-1,1]$ are
\begin{equation}
	\widetilde{\phi}_k^{\mathbf{r}}(m) = \begin{cases}
		\alpha_k \phi^{0}_k(m),
		& \quad 1 \leq m \leq L, \\
		\beta_k e^{-i\Delta} \phi^{0}_k(m),
		& \quad L+1 \leq m \leq 2L,
	\end{cases}
\end{equation}
and
\begin{equation}
	\widetilde{\phi}_k^{\mathbf{l}}(m) = \begin{cases}
		\alpha_k^* \phi^{0}_k(m),
		& \quad 1 \leq m \leq L, \\
		\beta_k^* e^{-i\Delta} \phi^{0}_k(m),
		& \quad L+1 \leq m \leq 2L,
	\end{cases}
\end{equation}
with the same eigenvalues and wavenumber as in \eqref{eq:defect_unitary_eigenvalue}. The prefactors $\alpha_k$ and $\beta_k$ satisfy the relation in \eqref{eq:prefactor_defect_wf}
together with the additional constraint
\be
\alpha_k\, \beta_k = \lambda,
\ee
which fixes the branch of the square root.
The scattering matrices constructed from the right and left eigenmodes are denoted by $S^{\mathbf r}$ and $S^{\mathbf l}$, respectively. We use the shorthand notation $S^{\mathbf r}\equiv S$ and $S^{\mathbf l}\equiv \bar S$. Here, the overbar in $\bar S$ is a notational label for the left-eigenmode scattering matrix and should not be interpreted as complex conjugation of $S$. The two matrices are defined explicitly as
\begin{equation}
	\begin{aligned}
		S^{\mathbf r}(k) &= S
		= \begin{pmatrix}
			\mu & e^{+i\Delta} \lambda \\
			e^{-i\Delta} \lambda & -\mu
		\end{pmatrix},
		\\
		S^{\mathbf l}(k) &= \bar{S}
		= \begin{pmatrix}
			\mu^* & e^{+i\Delta} \lambda^* \\
			e^{-i\Delta} \lambda^* & -\mu^*
		\end{pmatrix} ,
	\end{aligned}
\end{equation}
with $\det S = \det \bar S = -1$.
Moreover, by considering to multiply an imaginary unit, for $\lambda \in \mathbb{C} \setminus [-1,1]$,
\begin{equation}
	\hat{S}_{\rm non\text{-}unitary}^{\,{\rm complex}\,\lambda}=iS \in SL(2,\mathbb C),
\end{equation}
and similarly $\hat{\bar S}_{\rm non\text{-}unitary}^{\,{\rm complex}\,\lambda} = i\bar S \in SL(2,\mathbb C)$ for the left eigenmodes, with $\det\hat{S}=\det\hat{\bar S}=1$. In contrast to the Hermitian case, the scattering matrices are no longer unitary, and the transmission and reflection coefficients no longer sum to unity in general. This can be probed through a `` scattering experiment'' as in Ref.~\cite{2020_Meineri, 2025_Barad}. 

\smallskip

It is instructive to discuss separately the real non-Hermitian branches,
$\lambda\in(-\infty,-1)\cup(1,\infty)$. In
\eqref{eq:unitary_su2_scattering}, the unitary interface was shown to form an
$SU(2)$ family of scattering matrices, parametrized by
$\lambda=\cos\varphi$, where
$\varphi\in[-\frac{\pi}{2},\frac{\pi}{2}]$.
The extension to the real non-Hermitian branches can be viewed as the analytic continuation
\be
\varphi\to i\varphi,\quad \varphi\in \mathbb R,
\ee
which yields
\be
\begin{aligned}
	\hat{S}_{\rm non\text{-}unitary}^{\,{\rm real}\,\lambda}
	&= \begin{pmatrix}
		i \cosh \varphi & -e^{+i\Delta} \sinh\varphi \\
		-e^{-i\Delta} \sinh\varphi & -i\cosh\varphi
	\end{pmatrix} \in SU(1,1)
\end{aligned}
\ee
where $\hat S\equiv iS$.
In this regime, $\mu=i|\mu|$ is purely imaginary, and the usual unitary
transport sum rule is violated:
\be\label{eq:su11_transmission}
\mathcal{T}+\mathcal{R}
=\lambda^2+|\mu|^2
=2\lambda^2-1>1.
\ee
This indicates amplification of the probability due to the non-unitary nature. Furthermore, the ordinary unitarity condition $S^\dagger S=\Id$ is replaced by the pseudo-unitarity condition
\be
S^\dagger \sigma_z S=-\sigma_z.
\ee
Consequently the conserved quantity
\be
J_{\rm in} = |a^{{\rm I{\,,}in}}|^2 + |a^{{\rm II{\,,}in}}|^2,
\quad
J_{\rm out} = |a^{{\rm I{\,,}out}}|^2 + |a^{{\rm II{\,,}out}}|^2,
\ee
becomes pseudo current
\be
\widetilde{J}_{\rm in}
= - |a^{{\rm I{\,,}in}}|^2 + |a^{{\rm II{\,,}in}}|^2,
\quad
\widetilde{J}_{\rm out}
= |a^{{\rm I{\,,}out}}|^2 - |a^{{\rm II{\,,}out}}|^2 .
\ee
where the lattice modes are defined in \eqref{eq:relation_modes_wavefunction}.
Then the conservation law becomes $\widetilde{J}_{\rm in}=\widetilde{J}_{\rm out}$, which can be obtained by considering
\begin{equation}
	-a_{\rm in}^\dagger \sigma_z a_{\rm in} = a_{\rm in}^\dagger S^\dagger \sigma_z S a_{\rm in} = a_{\rm out}^\dagger \sigma_z a_{\rm out},
\end{equation}
where $a_{\rm in} = (a^{\rm I{, \,}in}, a^{\rm II{, \,}in})^T$ and $a_{\rm out} = (a^{\rm I{, \,}out}, a^{\rm II{, \,}out})^T$ with the modes defined in \eqref{eq:relation_modes_wavefunction}.

\subsection{Extension to interface on a circle}\label{sec:pbc_defect}

While the open-chain construction with a unitary interface is well studied, the exact realization of conformal interfaces on a periodic lattice is much less understood. The subtlety is not the local form of the interface, which can be chosen to be the same as in the open-chain case, but its global implementation on the circle. Inserting a single such interface into a periodic chain does not lead to an exactly solvable conformal defect: the single-particle spectrum and wavefunctions no longer admit simple closed forms. As discussed above, the similarity transformation is the key ingredient behind the exactly solvable open-chain construction. Here we show that the same idea selects a solvable periodic geometry, with two identical interfaces placed symmetrically on the circle.

We start from the defect-free periodic chain without any twist
\be
\hat{\mathbf H}^{\rm PBC,0}
= -\frac{1}{2}\sum_{m=1}^{2L}
\left(c_m^\dagger c_{m+1}+{\rm h.c.}\right),
\ee
with $c_{2L+1}\equiv c_1$, and denote its single-particle kernel by $H^{\rm PBC,0}$. Its eigenvalues are
\be\label{eq:energy_pbc}
E_k^{\rm PBC}
= -\cos \theta_k^{\rm PBC},
\quad
\theta_k^{\rm PBC}=\frac{2\pi k}{2L},
\ee
with $k=1,\cdots,2L$, and the corresponding eigenfunctions are
\be
\phi_k^{\rm PBC,0}(m)
= \frac{1}{\sqrt{2L}}\,
\exp\!\left(i m\theta_k^{\rm PBC}\right).
\ee
We then apply the similarity transformation
\be
\widetilde{H}^{\rm PBC}
= V H^{\rm PBC,0}\, V^{-1},
\ee
where $V$ is the matrix obtained in \eqref{eq:element_orth_matrix_defect}.
Although $V$ was derived from the open-chain wavefunctions, here we use its
action on the real-space couplings. The resulting nonzero matrix elements are
\begin{equation}
	\begin{aligned}
		\widetilde{H}^{\rm PBC}_{m,m+1}
		= \widetilde{H}^{\rm PBC}_{m+1,m}
		&=
		\begin{cases}
			-\frac{1}{2}, & m\neq L,\,2L, \\
			-\frac{\lambda}{2}, & m=L,\,2L,
		\end{cases}
		\\
		-\widetilde{H}^{\rm PBC}_{L,L}
		= \widetilde{H}^{\rm PBC}_{L+1,L+1}
		&=
		\widetilde{H}^{\rm PBC}_{2L,2L}
		= -\widetilde{H}^{\rm PBC}_{1,1}
		= \frac{\mu}{2},
	\end{aligned}
\end{equation}
where $m+1$ is understood modulo $2L$. Thus the circle contains two symmetric
interfaces, located at the bonds $(L,L+1)$ and $(2L,1)$, each with the same
local form as the open-chain interface. Since $\widetilde{H}^{\rm PBC}$ is
related to the defect-free Hamiltonian by a similarity transformation, its
spectrum remains exactly the same as in \eqref{eq:energy_pbc}.

For each nontrivial degenerate pair of momenta $k$ and $2L-k$, $E_k^{\rm PBC}=E_{2L-k}^{\rm PBC} = -\cos\frac{k\pi}{L}$,
one can choose two real standing-wave representatives of the same eigenspace, which gives
\be
\widetilde{\phi}_{k,+}^{\rm PBC}(m)
=
\begin{cases}
	\alpha_{+} \cos\!\left[\left(m-\frac12\right)\theta_k^{\rm PBC}\right],
	& 1 \leq m \leq L, \\
	\beta_{+} \cos\!\left[\left(m-\frac12\right)\theta_k^{\rm PBC}\right],
	& L+1 \leq m \leq 2L,
\end{cases}
\ee
and
\be
\widetilde{\phi}_{k,-}^{\rm PBC}(m)
=
\begin{cases}
	\alpha_{-} \sin\!\left[\left(m-\frac12\right)\theta_k^{\rm PBC}\right],
	& 1 \leq m \leq L, \\
	\beta_{-} \sin\!\left[\left(m-\frac12\right)\theta_k^{\rm PBC}\right],
	& L+1 \leq m \leq 2L,
\end{cases}
\ee
where
\be
\alpha_{\pm}^2 = 1 \pm \mu \,, \quad \beta_{\pm}^2 = 1 \mp \mu \,, \quad
\alpha_{\pm} \beta_{\pm} = \lambda \,.
\ee
The labels $+$ and $-$ denote the two independent
standing-wave combinations within the degenerate subspace. As in the
open-chain case, the local scattering analysis gives the same mode-independent
matrices $S$ and $\bar S$ in both the Hermitian and non-Hermitian regimes.

One can further fuse a $U(1)$ topological defect with each conformal interface, assigning phases $\Delta_1$ and $\Delta_2$. Unlike in the open-chain geometry, the total phase
\be
\Delta=\Delta_1+\Delta_2
\ee
cannot in general be gauged away and therefore affects the spectrum. For $\Delta=0$ modulo $2\pi$, the spectrum remains unchanged, although the eigenfunctions acquire additional phases. For nonzero total phase, the two defects generate an effective flux through the circle, shifting the momentum quantization to
\be\label{eq:energy_spectrum_pbc_flux}
E_k^{\rm circle}
= -\cos \theta_k^{\rm circle},
\quad
\theta_k^{\rm circle}(\Delta)
= \frac{2\pi k+\Delta_{\rm eff}}{2L},
\ee
with $k=1,\cdots,2L$ and
\be
\Delta_{\rm eff}
=
\arccos\!\left(
1-2\lambda^2\sin^2\frac{\Delta}{2}
\right).
\ee
Consequently, when the total phase is nonzero, the defected Hamiltonian is similar to a defect-free periodic chain with a $\lambda$-dependent effective twist. This is a global effect of the circle geometry. Nevertheless, each interface still supports mode-independent scattering matrix at the interface position. In particular, for the first interface on bond $(L,L+1)$, it is
\be
S_1
=
\begin{pmatrix}
	\mu & e^{+i\Delta_1}\lambda \\
	e^{-i\Delta_1}\lambda & -\mu
\end{pmatrix},
\ee
whereas for the second interface on bond $(2L,1)$, it is
\be
S_2
=
\begin{pmatrix}
	\mu & e^{-i\Delta_2}\lambda \\
	e^{+i\Delta_2}\lambda & -\mu
\end{pmatrix}.
\ee
The reversed phase convention in $S_2$ reflects the opposite orientation of the second interface, which interchanges the roles of the left- and right-moving modes.

\subsection{Correlation and entanglement: complex-valued effective central charge for non-unitary interfaces}

\begin{figure*}
\centering
\includegraphics[width=\textwidth]{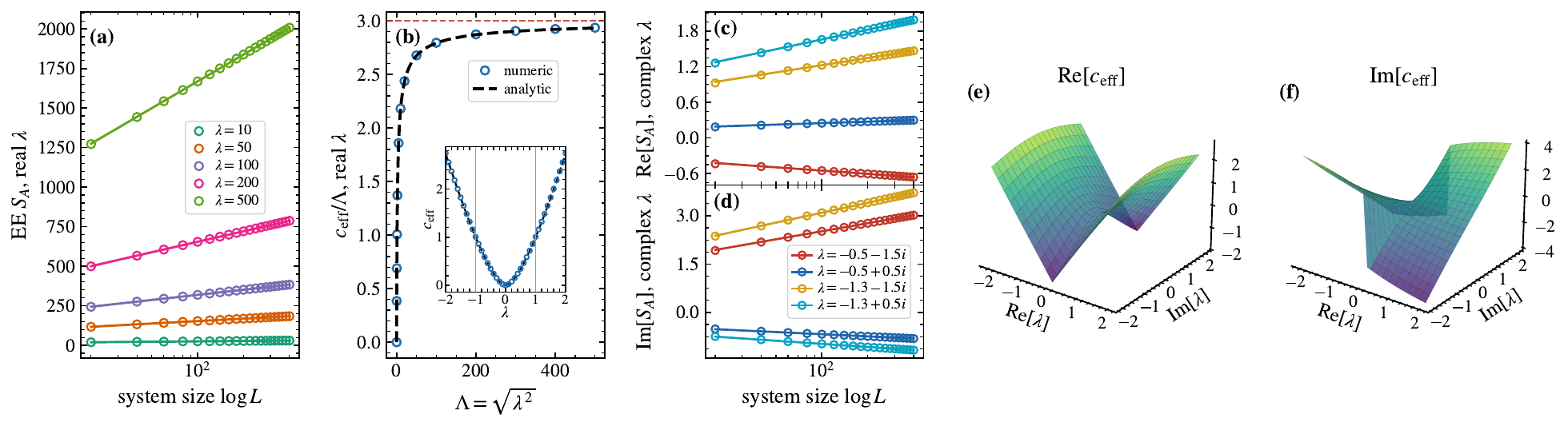}
\caption{Half-chain entanglement entropy (EE) $S_A$ and effective central charge $c_{\rm eff}$ for the defected free-fermion chain of length $2L$ defined in Eq.~\eqref{eq:defect_Ham}, with the entanglement cut placed at the interface. In panels (a), (c), and (d), open circles denote raw EE data obtained via lattice simulation for $L\in[20,400]$, and solid lines denote linear fits to $S_A=\frac{c_{\rm eff}}{6}\log L+b$ over $L\in[200,400]$. The $c_{\rm eff}$ values shown in panels (b), (e), and (f) are obtained by applying the same fitting procedure independently at each sampled interface parameter $\lambda$.
(a) $S_A$ versus $\log L$ for real $\lambda=10,50,100,200,500$, yielding $c_{\rm eff}=21.81, 133.90, 279.63, 574.93, 1468.32$, respectively, with relative deviations from Eq.~\eqref{eq:effective_central_charge} below $0.2\%$.
(b) $c_{\rm eff}/\Lambda$ versus $\Lambda=\sqrt{\lambda^2}=|\lambda|$ for real $\lambda$, compared with the analytic prediction in Eq.~\eqref{eq:effective_central_charge}, shown as the dashed curve. The red dashed line represents the large-$\Lambda$ asymptote $c_{\rm eff}/\Lambda=3$. Inset: $c_{\rm eff}$ versus $\lambda$, illustrating its evenness in $\lambda$ and continuity at $\lambda=0$ and $\lambda=\pm1$; solid lines at $\lambda=\pm1$ separate the unitary ($|\lambda|<1$) and non-unitary ($|\lambda|>1$) regimes.
(c,d) Real and imaginary parts of $S_A$ versus $\log L$ for complex couplings $\lambda=-0.5-1.5i, -0.5+0.5i, -1.3-1.5i, -1.3+0.5i$, with extracted $c_{\rm eff}=-0.476+2.160i, 0.216-0.580i, 1.047+2.667i, 1.427-0.840i$, respectively, agreeing with Eq.~\eqref{eq:effective_central_charge} within $\sim0.02$.
(e,f) Real and imaginary parts of $c_{\rm eff}$ for $\mathrm{Re}[\lambda],\mathrm{Im}[\lambda]\in[-2,2]$, sampled with grid spacing $0.1$ in both directions. A comparison with the analytic result in \eqref{eq:effective_central_charge} indicates an agreement within $\sim0.02$ over the domain shown.
}\label{fig:numeric_EE}
\end{figure*}

Since the interface Hamiltonians considered above have real single-particle
spectra, the half-filled many-body state is well defined by occupying the
negative-energy modes. In the non-Hermitian regimes, we use the corresponding
biorthogonal ground states, denoted by $\langle G_L|$ and $|G_R\rangle$. Since
the state is Gaussian, its reduced-state data are encoded in the correlation
matrix~\cite{Chung2001, Peschel2003, Bravyi2004, Eisler_2009_RDM, 2019_Chang}
\begin{equation}
	\widetilde{C}_{m,n}
	=
	\langle G_L | c_m^\dagger c_n | G_R \rangle
	=
	\sum_{\text{occupied } k}
	\left[\widetilde{\phi}^{\mathbf l}_k(m)\right]^*
	\widetilde{\phi}^{\mathbf r}_k(n).
\end{equation}
In the Hermitian case, $\langle G_L|=\langle G|$ and $|G_R\rangle=|G\rangle$.
Remarkably, the functional dependence on the interface parameter $\lambda$ is identical in the Hermitian and non-Hermitian regimes.
In particular, for the half subsystem
$A=[1,L]$, the restricted correlation matrix satisfies~\cite{Eisler_2010_EE_fermion_defect,Eisler_2012_defect_boson}
\begin{equation}
\label{eq:relation_CM_defect_free}
	\widetilde{C}_A(\Id-\widetilde{C}_A)
	=
	\lambda^2\, C^0_A(\Id-C^0_A),
\end{equation}
where $C^0_A$ is the correlation matrix of the defect-free chain ($\lambda=1$) and $\Id$ is the $L\times L$ identity matrix.

The EE for a subsystem $A$ is defined by
\be\label{eq:def_EE}
S_A=-\Tr(\rho_A\log\rho_A),
\ee
where $\rho_A$ is the reduced density matrix of $A$. In the
non-Hermitian case, $\rho_A$ is understood as the biorthogonal reduced density
matrix obtained from $|G_R\rangle\langle G_L|$ after tracing out $\bar A$. In general, calculating EE requires the spectrum of $\rho_A$, or equivalently the spectrum of entanglement Hamiltonian $H_E = -\log \rho_A$, known as \emph{entanglement spectrum} (ES). For a free fermion Gaussian state, the single-particle ES $\varepsilon_l$ is determined by the eigenvalues $\xi_l$ of $C_A$ through~\cite{Peschel2003,Eisler_2009_RDM}
\begin{equation}\label{eq:free_fermion_ES}
	\xi_l=\frac{1}{e^{\varepsilon_l}+1} ,
\end{equation}
and the EE is given by
\begin{equation}\label{eq:free_fermion_EE_correlation_matrix}
	S_A=\Tr[g(C_A)], \quad g(u)=-u\log u-(1-u)\log(1-u).
\end{equation}
From \eqref{eq:free_fermion_ES}, one has
\be
\xi_l(1-\xi_l)=\frac{1}{4\cosh^2(\varepsilon_l/2)}.
\ee
The relation \eqref{eq:relation_CM_defect_free} then gives
\begin{equation}\label{eq:relation_ES_defect}
	\cosh\frac{\widetilde{\varepsilon}_l}{2}
	=
	\frac{1}{\sqrt{\lambda^2}}
	\cosh\frac{\varepsilon_l^0}{2},
\end{equation}
where $\widetilde{\varepsilon}_l$ and $\varepsilon_l^0$ are the single-particle ES of the interface and defect-free ground states, respectively. We keep $\sqrt{\lambda^2}$ rather than $\lambda$ to make the evenness under $\lambda\to-\lambda$ explicit. The ES are real when $\lambda\in[-1,1]$, while generally become complex when $\lambda\in\mathbb C\setminus[-1,1]$.

In the continuum limit, the low-lying defect-free ES has a
linear dispersion. The entropy can then be written as \cite{Tang2024_critical_weak_measurement}
\begin{equation}
\label{eq:SA_lattice_integral}
	S_A
	=
	\int_{-\infty}^{+\infty} d\varepsilon^0\,
	g\!\left[\widetilde{\xi}(\varepsilon^0)\right]
	\int_{\epsilon}^{L_A-\epsilon} dx\, P_\xi(x),
\end{equation}
where $P_\xi(x)$ is the real-space density of entanglement-spectrum levels,
determined by the conformal map from the entanglement-cut geometry to an
annulus (we will discuss this map in the field theory description later).
The parameter $\epsilon$ in $\int_\epsilon^{L_A-\epsilon}$ serves as a short-distance cutoff \cite{2016_Cardy_Tonni}.
For an open chain with subsystem $A$ attached to one boundary, this
density gives
\begin{equation}
	\int_{\epsilon}^{L_A-\epsilon} dx\,P_\xi(x)
	=
	\frac{1}{2\pi^2}
	\log\frac{L}{\epsilon} + \mathcal{O}(1).
\end{equation}
Thus
\begin{equation}
\label{eq:SA_lattice}
	S_A
	=
	\frac{I(\lambda)}{2\pi^2}
	\log\frac{L}{\epsilon} + \mathcal{O}(1)
	=
	\frac{c_{\rm eff}}{6}
	\log\frac{L}{\epsilon} + \mathcal{O}(1),
\end{equation}
with
\begin{equation}
	I(\lambda)
	=
	\int_{-\infty}^{+\infty} d\varepsilon^0\,
	g\!\left[\widetilde{\xi}(\varepsilon^0)\right],
\end{equation}
and
\be
c_{\rm eff}=\frac{3}{\pi^2}I(\lambda).
\ee
Furthermore, the function $g(\widetilde\xi)$ in \eqref{eq:SA_lattice} can be rewritten as
\begin{equation}
	g(\widetilde{\xi})
	=
	\log\!\left(2\cosh\frac{\widetilde{\varepsilon}}{2}\right)
	-
	\frac{\widetilde{\varepsilon}}{2}
	\tanh\frac{\widetilde{\varepsilon}}{2},
\end{equation}
where $\widetilde{\varepsilon}$ is related to $\varepsilon^0$ through \eqref{eq:relation_ES_defect}.
Then the integral $I(\lambda)$ can be evaluated in a closed form as follows \cite{Eisler_2010_EE_fermion_defect}
\begin{equation}\label{eq:effective_central_charge}
	\begin{aligned}
		c_{\rm eff}
		&=
		-\frac{6}{\pi^2}
		\sum_{\kappa=\pm1}
		\Big[
		(1+\kappa \Lambda)\,
		{\rm Li}_2(-\kappa \Lambda)
		\\
		&\hspace{2cm}
		+
		(1+\kappa \Lambda)\,
		\log(1+\kappa \Lambda)\,
		\log \Lambda
		\Big],
	\end{aligned}
\end{equation}
with $\Lambda = \sqrt{\lambda^2}$. This closed form is generally complex-valued for $\lambda \in \mathbb{C} \setminus \mathbb{R}$, and the result is obviously even in the interface parameter, $c_{\rm eff}(-\lambda)=c_{\rm eff}(\lambda)$.
The analytical result is in excellent agreement with numerical calculations on the lattice, as shown in Fig.~\ref{fig:numeric_EE}.

It is particularly instructive to consider the case of real $\lambda$, for which the effective central charge remains real. Along the non-negative real axis, $c_{\rm eff}$ is continuous and monotonically increasing. In particular, its large-$\lambda$ asymptotic behavior follows directly from the integral
representation:
\begin{equation}
	c_{\rm eff}(\lambda)\sim 3\lambda,
	\qquad
	\lambda \gg 1.
\end{equation}
Thus the effective central charge grows without bound in the real non-Hermitian regime. This unbounded linear growth is the entanglement counterpart of the signal amplification encoded in the non-unitary scattering data for real $\lambda>1$, as discussed near \eqref{eq:su11_transmission}.

\section{CFT description of the non-unitary conformal interface}
\label{Sec:CFT}

The previous sections established the existence of exact conformal interfaces directly on the lattice. The key result is that the interface scattering matrix is exactly mode independent, rather than becoming so only in the low-energy limit. This holds in both the unitary and non-unitary regimes. Thus the lattice model is not merely a solvable regularization of a continuum interface. Instead, it realizes a family of exact interface fixed points and provides a controlled setting for developing the corresponding CFT description.

From the continuum perspective, one may formally impose complex gluing conditions for free fermions by choosing an appropriate scattering matrix. However, the origin and physical interpretation of such complex gluing data are, in general, unclear. In the present construction, the gluing data are uniquely determined by the exact scattering matrices derived from the lattice model. We shall use these data not only to construct the associated boundary and interface CFTs, but also to identify the proper framework for describing non-unitary boundaries and interfaces, in which the scattering data associated with the right and left eigenmodes must be treated independently.

\subsection{Construction of the boundary state}

We consider an interface located at $x=0$, separating two half-space free Dirac CFT: CFT$^{\rm I}$ on $x\in[-L,0)$ and CFT$^{\rm II}$ on $x\in(0,L]$. Folding along the interface converts it into a boundary condition of the product theory $\mathrm{CFT}^{\rm I}\otimes \overline{\mathrm{CFT}}{}^{\rm II}$ on $x\in[-L,0)$. We denote the Dirac fields in the two theories by $\Psi^{\rm I},\Psi^{\rm II}$, and their Dirac conjugates by $\overline{\Psi}^{\rm I},\overline{\Psi}^{\rm II}$, with $\overline{\Psi}\equiv\Psi^\dagger\gamma^0$.

The gluing conditions for $\Psi$ and $\overline{\Psi}$ are fixed by the scattering data at the interface. Let $\Psi_\pm=(\Psi_\pm^{\rm I},\Psi_\pm^{\rm II})^T$ denote the chiral and anti-chiral components, and similarly for $\overline{\Psi}_\pm$. At $x=0$, the interface imposes
\begin{equation}\label{eq:boundary_glue_variation}
	\Psi_+(0,\tau)=S^{-1}\Psi_-(0,\tau), \qquad \overline{\Psi}_+(0,\tau)=\overline{\Psi}_-(0,\tau)S .
\end{equation}
Through the mode expansion, these continuum fields describe the incoming and outgoing lattice modes discussed in the previous sections. The mode-independent scattering matrix obtained from the lattice model therefore becomes the gluing data of the CFT interface.

To define the corresponding boundary state in the closed-string channel, we rotate $w=x+i\tau$ to $w'=-iw=\tau-ix$, so that the interface lies on the real axis of the $w'$ plane. The gluing conditions become
\begin{equation}\label{eq:boundary_glue_variation_rot}
	\Psi_+(w')=iS^{-1}\Psi_-(w'), \qquad \overline{\Psi}_+(w')=i\overline{\Psi}_-(w')S ,
\end{equation}
valid at $\Im w'=0$. Equivalently, the boundary state $|b\rrangle$ is defined by requiring the corresponding operator insertions to vanish in closed-string amplitudes \cite{Cardy_1989},
\be
\begin{aligned}
	\lim_{\Im w'\to 0}\llangle a|\left[\Psi_+(w')-iS^{-1}\Psi_-(w')\right]|b\rrangle &=0,\\
	\lim_{\Im w'\to 0}\llangle a|\left[\overline{\Psi}_+(w')-i\overline{\Psi}_-(w')S\right]|b\rrangle &=0 .
\end{aligned}
\ee
Here $\llangle a|$ and $|b\rrangle$ denote the outgoing and incoming boundary states of the folded theory in the rotated channel. Since the theory is free, these linear constraints are solved by a Gaussian boundary state \cite{book2003_mirror_symmetry},
\begin{equation}\label{eq:incoming_boundary_state}
	|b\rrangle=\mathcal{N}_b\exp\!\left[i\sum_{n>0}\left(-\bar{\widetilde{\psi}}_{-n}S\psi_{-n}+\bar{\psi}_{-n}S^{-1}\widetilde{\psi}_{-n}\right)\right]|0\rangle .
\end{equation}
Here $\psi_n,\widetilde{\psi}_n,\bar{\psi}_n,\bar{\widetilde{\psi}}_n$ are the oscillator modes of $\Psi_+,\Psi_-,\overline{\Psi}_+,\overline{\Psi}_-$, respectively, each regarded as a two-component vector in the CFT$^{\rm I}$/CFT$^{\rm II}$ space. The expression above is written in the Neveu--Schwarz (NS) sector, where $n\in\mathbb{N}-\frac12$ takes positive half-integer values. The Ramond (R) sector can be treated within the same framework by including the standard zero-mode contribution. For simplicity, we do not discuss it explicitly here, since no additional formalism is required.

To obtain the outgoing boundary state associated with the same gluing condition, we introduce a reflection to the closed-string channel, $w=x+i\tau\mapsto \widetilde w=x-i\tau$, or equivalently $w'=\tau-ix\mapsto w''=-\tau-ix$. This reflection exchanges the chiral and anti-chiral components and converts the oscillator constraint defining $|b\rrangle$ into the corresponding constraint acting on the bra. The resulting state is
\begin{equation}\label{eq:outgoing_boundary_state}
	\llangle b|=\mathcal{N}_b^*\langle0|\exp\!\left[i\sum_{n>0}\left(-\bar{\psi}_{n}S^\dagger\widetilde{\psi}_{n}+\bar{\widetilde{\psi}}_{n}(S^{-1})^\dagger\psi_{n}\right)\right].
\end{equation}
Although $S^\dagger$ and $(S^{-1})^\dagger$ appear in this reflected state, no new scattering data have been introduced: both $|b\rrangle$ and $\llangle b|$ encode the same right-eigenmode scattering matrix $S$.

For a non-unitary interface, however, the Hamiltonian is non-Hermitian and the biorthogonal structure requires an additional input. The states $|b\rrangle$ and $\llangle b|$ are associated with the right-eigenmode scattering matrix $S$ only. A complete description must also include the boundary states associated with the left-eigenmode scattering matrix $\bar S$:
\begin{equation}
	\begin{aligned}
		|\bar b\rrangle&=\overline{\mathcal{N}}_b\exp\!\left[i\sum_{n>0}\left(-\bar{\widetilde{\psi}}_{-n}\bar S\psi_{-n}+\bar{\psi}_{-n}\bar S^{-1}\widetilde{\psi}_{-n}\right)\right]|0\rangle,\\
		\llangle\bar b|&=\overline{\mathcal{N}}_b^*\langle0|\exp\!\left[i\sum_{n>0}\left(-\bar{\psi}_{n}\bar S^\dagger\widetilde{\psi}_{n}+\bar{\widetilde{\psi}}_{n}(\bar S^{-1})^\dagger\psi_{n}\right)\right].
	\end{aligned}
\end{equation}
In the Hermitian case, the distinction between $S$ and $\bar S$ disappears. For non-unitary interfaces, the two sets of scattering data must be kept separately. As we will show in next section, the partition function naturally pairs the right- and left-eigenmode boundary states, making this biorthogonal structure essential to the non-unitary boundary/interface CFT formalism.

\subsection{Determination of partition function and boundary entropy}

After obtaining the explicit form of the boundary states describing the folded conformal interface, we now turn to the closed-string partition function. This determines the normalization factor $\mathcal N_b$ in the boundary state, which fixes the boundary entropy. More importantly, in the non-unitary case, it also clarifies the proper closed-string path integral and the role of the left- and right-eigenmode scattering data.

\subsubsection{\texorpdfstring{The free--$b$ and free--$\bar b$ channels}{The free-b and free-bar b channels}}

We begin with the open-chain geometry. In the unfolded theory, the two physical endpoints carry the same free boundary condition,
\be
\Psi_+^{j}(x_j,\tau)
=
-\Psi_-^{j}(x_j,\tau),
\qquad
\overline{\Psi}_+^{j}(x_j,\tau)
=
-\overline{\Psi}_-^{j}(x_j,\tau),
\ee
where $j={\rm I},{\rm II}$ labels the two CFTs separated by the interface at $x=0$, with $x_{\rm I}=-L$ and $x_{\rm II}=L$. After folding, the product theory $\mathrm{CFT}^{\rm I}
\otimes \overline{\mathrm{CFT}^{\rm II}}$
is defined on the strip $-L\leq x\leq0$. The boundary at $x=-L$ is the folded image of the two free endpoints, while the boundary at $x=0$ represents the interface.

The scattering matrix of the free boundary is $S_{\rm free}=-\Id$, and the corresponding boundary states are
\begin{equation}\label{eq:free_boundary_state}
	\begin{aligned}
		|\mathrm{free}\rrangle
		&=
		\exp\!\left[
		i\sum_{n>0}
		\left(
		\bar{\widetilde{\psi}}_{-n}\psi_{-n}
		-
		\bar{\psi}_{-n}\widetilde{\psi}_{-n}
		\right)
		\right]|0\rangle,
		\\
		\llangle\mathrm{free}|
		&=
		\langle0|
		\exp\!\left[
		i\sum_{n>0}
		\left(
		\bar{\psi}_{n}\widetilde{\psi}_{n}
		-
		\bar{\widetilde{\psi}}_{n}\psi_{n}
		\right)
		\right].
	\end{aligned}
\end{equation}
We work in a fixed fermionic spin structure. In the NS sector, $n\in\mathbb{N}-\frac12$ takes positive half integers. The normalization $\mathcal N_{\rm free}=1$ is fixed by open--closed consistency.

The closed-channel amplitude between the free boundary and the interface boundary $b$ is
\be\label{eq:partition_free_b}
Z_{{\rm free},b}
=
\llangle\mathrm{free}|e^{-L H_{\rm c}}|b\rrangle,
\qquad
H_{\rm c}
=
\frac{2\pi}{\beta}
\sum_{j={\rm I},{\rm II}}
\widehat{H}^j \,,
\ee
where $\widehat{H}^j=L_0^j+\bar L_0^j-\frac{c_j}{12}$
Here $L$ is the width of the folded strip and $\beta$ is its circumference. We take $c_{\rm I}=c_{\rm II}=c$ and introduce the closed-channel modular parameter
\be
q=e^{-4\pi L/\beta}=e^{2\pi i\tau},
\qquad
\tau=\frac{2iL}{\beta}.
\ee
With this convention, each oscillator contraction at level $n$ contributes $q^n$, while the folded vacuum contributes $q^{-2c/24}$. Substituting Eqs.~\eqref{eq:incoming_boundary_state} and \eqref{eq:free_boundary_state} into \eqref{eq:partition_free_b} gives
\be
Z_{{\rm free},b}
=
\mathcal N_b\, q^{-\frac{2c}{24}}
\prod_{n>0}
\det\!\left(\Id+q^nS\right)
\det\!\left(\Id+q^nS^{-1}\right).
\ee

For the interface family considered here,
\be
S=
\begin{pmatrix}
	\mu e^{i\Gamma} & \lambda e^{i\Delta}\\
	\lambda e^{-i\Delta} & -\mu e^{-i\Gamma}
\end{pmatrix},
\qquad
\mu^2+\lambda^2=1,
\ee
with the associated symmetry structure $\hat S=iS\in SU(2)$.
A direct calculation yields
\be
Z_{{\rm free},b}
=
\mathcal N_b \, q^{-\frac{2c}{24}}
\prod_{n>0}
\left[
(1-q^{2n})^2
+
4q^{2n}\mu^2\sin^2\Gamma
\right].
\ee
The transmission phase $\Delta$ drops out, which is consistent with the common knowledge that a $U(1)$ transmission phase does not change the open-string spectrum of a chain with open ends. While the reflection phase $\Gamma$ generally affects the amplitude, for the interfaces studied in this work, $\Gamma=0$, and therefore
\begin{equation}\label{eq:partition_free_b_gamma_zero}
	Z_{{\rm free},b}
	=
	\mathcal N_b \, q^{-\frac{2c}{24}}
	\prod_{n>0}(1-q^{2n})^2
	=
	\mathcal N_b \, Z_{{\rm free},{\rm free}},
\end{equation}
where $Z_{{\rm free},{\rm free}}$ is the defect-free partition function with free ends.

The normalization is fixed in the open channel. In the corresponding lattice model, the Hamiltonian containing the interface is related to the defect-free Hamiltonian by an invertible similarity transformation.
Consequently,
\be
Z^{\rm o}_{{\rm free},b}
=
\Tr e^{-\beta H_{\rm o}}
=
Z^{\rm o}_{{\rm free},{\rm free}}.
\ee
Together with \eqref{eq:partition_free_b_gamma_zero}, open--closed consistency fixes $\mathcal N_b=1$.
Recall that the boundary entropy \cite{1991_Affleck_Ludwig} is defined as $s_b=\log g_b \,, g_b=\langle0|b\rrangle=\mathcal N_b$, which gives
\begin{equation}
	g_b=1, \quad s_b=0
\end{equation}
for both the unitary and non-unitary interfaces studied here.
The same argument applies to the dual boundary state $\bar b$, constructed from the left-eigenmode scattering data, giving $g_{\bar b}=1, s_{\bar b}=0$.

In the Hermitian regime, $s_b=0$ throughout the interface family. Its value is fixed by the boundary-state normalization, while its independence of the interface coupling is consistent with the exact marginality of the corresponding interface operator, which generates an $SU(2)$ family of unitary conformal interfaces.
The non-Hermitian continuation is less studied. Previous work on the free boson argued that the $SL(2,\mathbb{C})$ boundaries is distinct from its $SU(2)$ subgroup, from the analyticity of the conformal manifold~\cite{Gaberdiel2001_SL2C_boundary,
Hasselfield2005_sineGordon_SL2C_boundary}.
In the present model, the exact similarity transformation preserves the open-channel spectrum, and the resulting $g$-factor remains $g_b=1$, regardless the interface is unitary or non-unitary. Thus, the partition function and boundary entropy alone do not distinguish the non-unitary interfaces from their unitary counterparts. This does not imply that the non-unitary continuation is physically trivial: these interfaces exhibit amplification in their scattering probabilities and a generally complex effective central charge extracted from entanglement scaling, both of which are intrinsic signatures of non-unitarity. Moreover, although for unitary theories one has the well-established monotonicity of the boundary entropy along RG flows~\cite{Friedan_2003}, in the present non-unitary case, the fixed-point result $s_b=0$ does not determine the behavior under perturbations away from this conformal family, see Appendix~\ref{app:RG} for a discussion from holopgraphic point of view. The corresponding non-unitary boundary RG flows must be studied separately. The Gaussian theory considered here provides a minimal and analytically controlled setting for such an investigation, which we leave to future work.

\subsubsection{\texorpdfstring{The $b$--$b$ and $\bar b$--$b$ channels}{The b-b and bar b-b channels}}

The free--$b$ and free--$\bar b$ amplitudes determine the individual boundary-state normalizations, but not the pairing of right- and left-eigenmode data. We therefore turn to amplitudes with interface boundary conditions at both ends of the folded cylinder, corresponding to the periodic lattice construction in Sec.~\ref{sec:pbc_defect}.

We first consider the $b$--$b$ channel, in which incoming and outgoing boundaries are constructed from the same right-eigenmode boundary data. Setting the transmission phases to zero for simplicity and using $\mathcal N_b=1$, we find
\begin{equation}
	\begin{aligned}
		Z_{bb}
		&=
		\llangle b|e^{-L H_{\rm c}}|b\rrangle
		\\
		&=
		q^{-\frac{2c}{24}}
		\prod_{n>0}
		\det\!\left[
		\Id-q^nS^\dagger S
		\right]
		\det\!\left[
		\Id-q^n(S^{-1})^\dagger S^{-1}
		\right].
	\end{aligned}
\end{equation}
For a unitary interface, $S^\dagger S=\Id$, and hence
\begin{equation}\label{eq:partition_func_bb_unitary}
	Z_{bb}^{\rm unitary}
	=
	q^{-\frac{2c}{24}}
	\prod_{n>0}(1-q^n)^4
	=
	Z_P,
\end{equation}
where $Z_P$ is the partition function of the folded defect-free Dirac theory on a circle, within the chosen spin structure.

For a non-unitary interface, $S^\dagger S\neq\Id$. On the phase-free branch, $S^{-1}=S$, and the same overlap becomes
\be
Z_{bb}
=
q^{-\frac{2c}{24}}
\prod_{n>0}
\left[
1
-
2q^n\left(
|\lambda|^2+|\mu|^2
\right)
+
q^{2n}
\right]^2.
\ee
This amplitude depends on the non-unitary interface parameter and therefore does not reproduce the similarity-invariant partition function of the periodic lattice model. Instead, it is more like a norm-type overlap sensitive to the nonorthogonality of the right eigenmodes.

The appropriate partition function is instead obtained in the $\bar b$--$b$ channel. Here $\bar S$ denote the left-eigenmode scattering matrix, which satisfies $\bar S^\dagger S=\Id$ as demonstrated in the lattice calculation.
Using also $\mathcal N_{\bar b}=\mathcal N_b=1$, we obtain
\begin{equation}\label{eq:partition_func_bar_b_b}
	\begin{aligned}
		Z_{\bar b b}
		&=
		\llangle\bar b|e^{-L H_{\rm c}}|b\rrangle
		\\
		&=
		q^{-\frac{2c}{24}}
		\prod_{n>0}
		\det\!\left[
		\Id-q^n\bar S^\dagger S
		\right]
		\det\!\left[
		\Id-q^n(\bar S^{-1})^\dagger S^{-1}
		\right]
		\\
		&=
		q^{-\frac{2c}{24}}
		\prod_{n>0}(1-q^n)^4
		=
		Z_P.
	\end{aligned}
\end{equation}
Thus the periodic lattice partition function is reproduced by the biorthogonal $\bar b$--$b$ amplitude, rather than by the $b$--$b$ overlap.

We may further fuse the outgoing and incoming boundaries with $U(1)$ topological defects carrying transmission phases $\Delta_1$ and $\Delta_2$. Let $\bar a\equiv\bar b(\Delta_1)$, $b\equiv b(\Delta_2)$.
With the orientation convention used in this work, the total inserted phase is $\Delta=\Delta_1+\Delta_2$. The corresponding partition function is
\begin{equation}
	\begin{aligned}
		Z_{\bar a b}
		&=
		\llangle\bar a|e^{-L H_{\rm c}}|b\rrangle
		\\
		&=
		q^{-\frac{2c}{24}}
		\prod_{n>0}
		\left[
		1
		-
		2q^n\left(
		\mu^2+\lambda^2\cos\Delta
		\right)
		+
		q^{2n}
		\right]^2,
	\end{aligned}
\end{equation}
with an effective twist
\be
\cos\Delta_{\rm eff}
=
\mu^2+\lambda^2\cos\Delta
=
1-2\lambda^2\sin^2\frac{\Delta}{2}.
\ee
This is identical with the lattice calculation in \eqref{eq:energy_spectrum_pbc_flux}, and therefore produces the standard flux-twisted free-fermion form.
For $\Delta=0$, this reduces to \eqref{eq:partition_func_bar_b_b}. In the fully transmitting limit, $\mu=0$ and $\lambda^2=1$, one has $\cos\Delta_{\rm eff}=\cos\Delta$; choosing the branch continuously connected to $\Delta=0$ gives $\Delta_{\rm eff}=\Delta$.

The consistency of $Z_{\bar b b}=Z_P$ indicates that, for the non-unitary interface theory, the incoming and outgoing boundary states should be associated with the scattering data of right and left eigenmodes,
\be
S\equiv S^{\mathbf r},
\qquad
\bar S\equiv S^{\mathbf l}.
\ee
This construction of using both right and left eigenstates is suggested by the lattice construction, but it also follows naturally from open--closed duality. In the open-string channel, a non-Hermitian Hamiltonian is traced in a biorthogonal basis. Since the closed-channel Hamiltonian $H_{\rm c}$ is the bulk unitary Hamiltonian of the folded theory, the left- and right-eigenmode data must be supplied by the boundary states. The appropriate closed-string partition function is therefore
\be
Z_{\bar b b}
=
\llangle\bar b|e^{-L H_{\rm c}}|b\rrangle
=
\Tr e^{-\beta H_{\rm o}},
\ee
rather than the $b$-$b$ overlap. Here, the left-right pairing selects the physical closed-string channel amplitude that is dual to the open-string partition function from energy spectrum.

\subsection{Entanglement entropy and interface operator in the non-unitary interface CFT}
\label{Sec:EE}

In the lattice calculation, we considered an open chain bipartitioned at the conformal interface and found a logarithmic entanglement entropy governed by a generally complex-valued effective central charge in \eqref{eq:SA_lattice}. We now reproduce this result directly in the unfolded interface CFT, and present the explicit form of the non-unitary interface operators in the original unfolded theory. The two CFTs occupy $x\in[-L,0)$ and $x\in(0,L]$, with the interface and the entangling point both located at $x=0$. To extract the universal scaling term, we introduce a UV cutoff $\epsilon$ around the entangling point \cite{2016_Cardy_Tonni}, and an IR cutoff $|w|=L$, where $w=x+i\tau$.

For computing EE, we adopt the \emph{replica trick} \cite{1994_Wilczek,2004_Cardy_Calabrese},
\begin{equation}
	S_A = -\left.
	\frac{\partial}{\partial n}
	\log \left[ \Tr \rho_A^n \right]
	\right|_{n=1}
	= -\left.
	\frac{\partial}{\partial n}
	\log Z_A^{(n)}
	\right|_{n=1}.
	\label{eq:replica_trick}
\end{equation}
where $\rho_A$ is the biorthogonal reduced density matrix of subsystem $A$ and
\begin{equation}
	Z_A^{(n)} \coloneqq \Tr\rho_A^n
	=
	\frac{\mathcal Z_n(A)}{[\mathcal Z_1(A)]^n}
\end{equation}
is the normalized replicated partition function. For $A=(-L,0)$, tracing out its complementary $\bar A=(0,L)$ leaves a branch cut along $A$ at $\tau=0$. Cyclically sewing $n$ copies along this cut produces the $n$-sheeted replica manifold.

Under the logarithmic map
\be
z=\log\frac{w}{\epsilon},
\ee
the replica manifold becomes an annulus,
\be
0\leq\Re z\leq W,
\quad
0\leq\Im z<2\pi n,
\quad
W=\log\frac{L}{\epsilon}.
\ee
The interface is mapped to the $2n$ lines \cite{2008_Sakai}
\be
\Im z=\frac{(2k-1)\pi}{2},
\qquad
k=1,\ldots,2n.
\ee
Time evolution along $\Im z$-direction within $\mathrm{CFT}^i$, $i={\rm I},{\rm II}$, is
generated by
\be
H_i^{\rm cyl}
=
\frac{2\pi}{W}\widehat H_i,
\qquad
\widehat H_i
=
L_0^{(i)}+\bar L_0^{(i)}-\frac{c_i}{12}.
\ee
Since neighboring interface lines are separated by a distance $\pi$ along $\Im z$-direction, it is convenient to introduce the modular parameter
\be
\widehat q
=
\exp\left(-\frac{2\pi^2}{W}\right).
\ee
The unnormalized replica partition function is then
\begin{equation}\label{eq:trace_replica_partition_func}
	\mathcal Z_n
	=
	\Tr_{\mathcal H_{\rm I}}
	\left(
	\mathcal I_{12}\,
	\widehat q^{\,\widehat H_{\rm II}}\,
	\mathcal I_{21}\,
	\widehat q^{\,\widehat H_{\rm I}}
	\right)^n,
\end{equation}
where
$\mathcal I_{12}:\mathcal H_{\rm II}\to\mathcal H_{\rm I}$
and
$\mathcal I_{21}:\mathcal H_{\rm I}\to\mathcal H_{\rm II}$
are the interface operators with opposite orientations.
Moreover, $\mathcal H_j, \, j = {\rm I}, {\rm II}$ denote the Hilbert space of the corresponding CFT$^j$ defined on a fixed-$\Im z$ slice of the annulus.

\subsubsection{The explicit form of interface operators}

The interface operators are obtained by unfolding the boundary states constructed in previous sections. In the NS sector, $\mathcal I_{12}$ factorizes over positive half-integer modes,
\begin{equation}\label{eq:I12_NS}
	\mathcal I_{12}^{\rm NS}
	=
	\left[
	\prod_{r\in\mathbb N-\frac12}
	\mathcal I_{12}(r)
	\right]
	P_{12}^{\rm NS} \,,
\end{equation}
where
\begin{equation}
	P_{12}^{\rm NS}
	=
	|0,{\rm NS}\rangle_{\rm I}\,
	{}_{\rm II}\langle0,{\rm NS}|
\end{equation}
is the map from the NS-sector ground state of CFT$^{\rm II}$ to that of CFT$^{\rm I}$. In the R sector, there is an additional zero-mode contribution, but it does not affect the leading scaling behavior of the EE.
The mode-resolved interface operator is
\begin{equation}\label{eq:I12_mode}
	\begin{aligned}
		\mathcal I_{12}(r)
		=
		\exp\Bigg\{
		i\Bigg[
		&-
		\begin{pmatrix}
			\bar{\widetilde\psi}_{-r}^{\rm I}
			&
			\bar\psi_r^{\rm II}
		\end{pmatrix}
		\bar D
		\begin{pmatrix}
			\psi_{-r}^{\rm I}\\[0.3ex]
			\widetilde\psi_r^{\rm II}
		\end{pmatrix}
		\\
		&+
		\begin{pmatrix}
			\bar\psi_{-r}^{\rm I}
			&
			\bar{\widetilde\psi}_r^{\rm II}
		\end{pmatrix}
		D^{-1}
		\begin{pmatrix}
			\widetilde\psi_{-r}^{\rm I}\\[0.3ex]
			\psi_r^{\rm II}
		\end{pmatrix}
		\Bigg]
		\Bigg\},
	\end{aligned}
\end{equation}
where
\be
D=
\begin{pmatrix}
	S_{11} & iS_{12}\\
	iS_{21} & -S_{22}
\end{pmatrix},
\qquad
\bar D=
\begin{pmatrix}
	S_{11} & -iS_{12}\\
	-iS_{21} & -S_{22}
\end{pmatrix}.
\ee
The coefficient difference between the interface operator \eqref{eq:I12_mode} and the folded boundary state \eqref{eq:incoming_boundary_state} arises from the exchange of chiral and anti-chiral orientations upon unfolding, but they represent the same physical gluing condition.

For each mode $r$, the Hilbert space factorizes into two four-dimensional fermionic sectors, which are spanned by the following bases
\begin{equation}\label{eq:interface_operator_basis_p}
	\begin{aligned}
		\mathcal B_{\rm I}^{(+)} & = \left( |0\rangle_{\rm I} , \bar{\widetilde{\psi}}{}_{-r}^{\rm I} \psi_{-r}^{\rm I} |0\rangle_{\rm I} , \psi_{-r}^{\rm I} |0\rangle_{\rm I} ,  \bar{\widetilde{\psi}}{}_{-r}^{\rm I} |0\rangle_{\rm I}  \right),
		\\
		\mathcal B_{\rm II}^{(+)*} & = \left(
		{}_{\rm II}\langle0| \,,\, {}_{\rm II}\langle0| \bar{\psi}_r^{\rm II} \widetilde{\psi}_r^{\rm II} \,,\, {}_{\rm II}\langle0| \bar\psi_{r}^{\rm II} \,,\, {}_{\rm II}\langle0| \widetilde{\psi}_{r}^{\rm II} \right) ,
	\end{aligned}
\end{equation}
and
\begin{equation}\label{eq:interface_operator_basis_m}
	\begin{aligned}
		\mathcal B_{\rm I}^{(-)} & = \left(
		|0\rangle_{\rm I} , \bar{\psi}{}_{-r}^{\rm I} \widetilde{\psi}_{-r}^{\rm I} |0\rangle_{\rm I} , \bar{\psi}_{-r}^{\rm I} |0\rangle_{\rm I} ,  {\widetilde{\psi}}{}_{-r}^{\rm I} |0\rangle_{\rm I}  \right),
		\\
		\mathcal B_{\rm II}^{(-)*} & = \left(
		{}_{\rm II}\langle0| \,,\, {}_{\rm II}\langle0| \bar{\widetilde{\psi}}_r^{\rm II} {\psi}_r^{\rm II} \,,\, {}_{\rm II}\langle0| \psi_{r}^{\rm II} \,,\, {}_{\rm II}\langle0| \bar{\widetilde{\psi}}{}_{r}^{\rm II}  \right).
	\end{aligned}
\end{equation}
In these bases, the matrix representation of the operator $\mathcal{I}_{12}(r)$ mapping from $\mathcal{H}_{\rm II} \to \mathcal{H}_{\rm I}$ is given by the matrix elements $I_{12}(r)_{ij} = {}_{\rm I}\langle i | \mathcal{I}_{12}(r) | j \rangle_{\rm II}$, which factorizes as:
\begin{equation}
	I_{12}(r)=I_{12}^{(+)}\otimes I_{12}^{(-)}
\end{equation}
and the block components $I_{12}^{(+)}$ and $I_{12}^{(-)}$ take the following explicit forms:
\begin{equation}\label{eq:I12_matrix}
\begin{aligned}
	I_{12}^{(+)}
	& =
	\begin{pmatrix}
		1 & -iS_{11} & 0 & 0\\
		iS_{22} & \det S & 0 & 0\\
		0 & 0 & -S_{21} & 0\\
		0 & 0 & 0 & S_{12}
	\end{pmatrix},
	\\
	I_{12}^{(-)}
	& =
	\begin{pmatrix}
		1 & i(S^{-1})_{11} & 0 & 0\\
		-i(S^{-1})_{22} & \det(S^{-1}) & 0 & 0\\
		0 & 0 & -(S^{-1})_{12} & 0\\
		0 & 0 & 0 & (S^{-1})_{21}
	\end{pmatrix},
\end{aligned}
\end{equation}
are the components in the two sectors with bases defined in \eqref{eq:interface_operator_basis_p} and \eqref{eq:interface_operator_basis_m}, respectively.

Reversing the interface orientation exchanges the two CFTs and replaces $S$ by $S^{-1}$, resulting in
\begin{equation}\label{eq:I21_from_I12}
	\mathcal I_{21}^{\rm NS}
	=
	\left.
	\mathcal I_{12}^{\rm NS}
	\right|_{
		S\leftrightarrow S^{-1},
		\ {\rm I}\leftrightarrow{\rm II}
	}.
\end{equation}
The matrix representation of the mode-resolved interface operator $\mathcal{I}_{21}(r)$ (which now maps from $\mathcal{H}_{\rm I} \to \mathcal{H}_{\rm II}$) similarly factorizes into the two sectors as:
\begin{equation}
I_{21}(r)=I_{21}^{(+)}\otimes I_{21}^{(-)},
\end{equation}
with
\begin{equation}\label{eq:I21_matrix}
\begin{aligned}
	I_{21}^{(+)}
	& =
	\begin{pmatrix}
		1 & -i(S^{-1})_{22} & 0 & 0\\
		i(S^{-1})_{11} & \det(S^{-1}) & 0 & 0\\
		0 & 0 & -(S^{-1})_{12} & 0\\
		0 & 0 & 0 & (S^{-1})_{21}
	\end{pmatrix},
	\\
	I_{21}^{(-)}
	& =
	\begin{pmatrix}
		1 & iS_{22} & 0 & 0\\
		-iS_{11} & \det S & 0 & 0\\
		0 & 0 & -S_{21} & 0\\
		0 & 0 & 0 & S_{12}
	\end{pmatrix}.
\end{aligned}
\end{equation}
In the non-unitary regime, the biorthogonality relation
$S^{-1}=\bar S^\dagger$ shows that the two orientations naturally incorporate the
right- and left-eigenmode scattering data identified in the previous
subsection.

\subsubsection{Replica partition function and entanglement entropy}

The oscillator propagator in each four-dimensional sector is
\be
P(r)
=
\operatorname{diag}
\left(
1,\widehat q^{\,2r},
\widehat q^{\,r},
\widehat q^{\,r}
\right).
\ee
We therefore define the two mode-resolved transfer matrices
\be
T_\sigma(r)
=
I_{12}^{(\sigma)}
P(r)
I_{21}^{(\sigma)}
P(r),
\qquad
\sigma=\pm.
\ee
Separating out the vacuum contribution, the replica partition function
factorizes as
\begin{equation}\label{eq:Zn_mode_factorization}
	\mathcal Z_n
	=
	\widehat q^{\,2nE_0}
	\prod_{r\in\mathbb{N}-\frac12}
	\Tr\!\left[T_+(r)^n\right]
	\Tr\!\left[T_-(r)^n\right],
\end{equation}
where $E_0=-c/12=-1/12$ is the vacuum eigenvalue of $\widehat H=\widehat{H}^{\rm I}=\widehat{H}^{\rm II}$ for either CFT.

For the interfaces considered here, the reflection phase $\Gamma=0$, and the transmission phase $\Delta$ drops out from the eigenvalues of $T_{\pm}(r)$. For odd positive integers $n$, a direct calculation gives
\begin{equation}\label{eq:mode_trace_replica}
	\Tr\!\left[T_\sigma(r)^n\right]
	=
	\prod_{k=1}^{n}
	\left[
	1
	+
	2\cos(2\nu_k) \hat{q}^{\,r}
	+
	\hat{q}^{\,2r}
	\right],
\end{equation}
where
\begin{equation}\label{eq:replica_effective_flux}
	\cos(2\nu_k)
	=
	1-2\lambda^2\sin^2\frac{\pi k}{n},
	\qquad
	k=1,\ldots,n.
\end{equation}
The angles $2\nu_k$ may be viewed as effective fluxes in the replica-momentum sectors. For even $n$, the mode trace contains an additional fermion-parity contribution. As discussed in Ref.~\cite{Brehm2015_EE_interfece_ising}, here the analytic calculation of $n$ is safe for obtaining the entanglement entropy at the replica limit $n=1$.

Since the two fermionic sectors give identical contributions,
Eqs.~\eqref{eq:Zn_mode_factorization} and
\eqref{eq:mode_trace_replica} yield
\begin{equation}\label{eq:unnormalized_replica_partition}
	\mathcal Z_n
	=
	\widehat q^{\,2nE_0}
	\prod_{k=1}^{n}
	\prod_{r\in\mathbb{N}-\frac12}
	\left[
	1
	+
	2\cos(2\nu_k)\widehat q^{\,r}
	+
	\widehat q^{\,2r}
	\right]^2.
\end{equation}
At $n=1$, $\nu_1=0$, and hence
\be
\mathcal Z_1
=
\widehat q^{\,2E_0}
\prod_{r\in\mathbb{N}-\frac12}
\left(
1+\widehat q^{\,r}
\right)^4.
\ee
The normalized replica partition function is therefore
\begin{equation}\label{eq:replica_partition_theta}
\begin{aligned}
	Z_A^{(n)}
	& =
	\prod_{k=1}^{n}
	\prod_{r\in\mathbb{N}-\frac12}
	\left[
	\frac{
		1+2\cos(2\nu_k)\widehat q^{\,r}
		+\widehat q^{\,2r}
	}{
		(1+\widehat q^{\,r})^2
	}
	\right]^2
	\\ & =
	\prod_{k=1}^{n}
	\left[
	\frac{
		\theta_3(\nu_k|\tau)
	}{
		\theta_3(0|\tau)
	}
	\right]^2 ,
\end{aligned}
\end{equation}
where $\theta$ is the standard Jacobi-$\theta$ function and we defined the modular parameter $\tau$ by $\hat{q}=\exp(i\pi\tau)=\exp(-2\pi^2/W)$, with $W=\log \frac{L}{\epsilon}$.

Using the modular transformation of the Jacobi-$\theta$ function, one finds that in the large-system-size limit, i.e., $W \gg 1$ and $t=-i\tau \ll 1$, the asymptotic behavior of \eqref{eq:replica_partition_theta} is \cite{2008_Sakai,Brehm2015_EE_interfece_ising}
\be
\log Z_A^{(n)}
\sim
-\frac{W}{\pi^2}f(n) \,,
\ee
where
\begin{equation}
f(n)
=
\sum_{k=1}^{n}\nu_k^2 = \sum_{k=1}^{n}\arcsin^2\left(
\Lambda\sin\frac{\pi k}{n}
\right) ,
\end{equation}
with $\Lambda = \sqrt{\lambda^2}$.
This asymptotic behavior is valid not only for $\lambda \in [-1,1]$ but also for general $\lambda \in \mathbb{C}$.
The entanglement entropy therefore takes the form
\begin{equation}\label{eq:EE_from_replica}
\begin{aligned}
	S_A
	= - \left. \partial_n \log Z_A^{(n)} \right|_{n=1}
	& \sim \frac{W}{\pi^2}
	\left.
	\frac{\partial f(n)}{\partial n}
	\right|_{n=1}
	\\ & \quad =
	\frac{c_{\rm eff}}{6}
	\log\frac{L}{\epsilon}
	+O(1) \,,
\end{aligned}
\end{equation}
where
\be
c_{\rm eff}
=
\frac{6}{\pi^2}
\left.
\frac{\partial f(n)}{\partial n}
\right|_{n=1}.
\ee
The replica limit $\lim_{n\to1} \partial_n$ for $f(n)$ can be evaluated using the Abel--Plana formula, giving
\begin{equation}\label{eq:replica_limit_integral}
	\left.
	\frac{\partial f(n)}{\partial n}
	\right|_{n=1}
	=
	4\int_0^\infty
	\frac{u\,du}{
		\exp\!\left[
		2\,\operatorname{arcsinh}
		\left(
		\frac{\sinh u}{\Lambda}
		\right)
		\right]-1
	} .
\end{equation}
Evaluating the integral yields
\begin{equation}\label{eq:replica_limit_closed_form}
	\begin{aligned}
		\left.
		\frac{\partial f(n)}{\partial n}
		\right|_{n=1}
		=
		-\sum_{\kappa=\pm1}
		(1+\kappa\Lambda)
		\Big[
		&\operatorname{Li}_2(-\kappa\Lambda)
		\\
		&+
		\log(1+\kappa\Lambda)\log\Lambda
		\Big].
	\end{aligned}
\end{equation}
This leads to the same closed form of the effective central charge as in \eqref{eq:effective_central_charge} obtained from lattice calculation.

\section{Global quenches from non-unitary boundary states}
\label{Sec:Quench}

So far we have studied non-unitary boundary conditions in unitary CFTs from the open-string point of view, i.e. we have placed the boundaries in the spatial direction.
Given the existence of such non-unitary boundary conditions, in this section, we would like to consider an application of them from the closed string point of view. In other words, we will put boundaries on the Euclidean temporal direction. Instead of focusing on the free CFTs as in the previous sections, we will study universal features in generic CFTs.

The setup we would like to consider is that of a global quantum quench. Using standard unitary conformal boundary conditions \cite{Cardy_1989}, Cardy and Calabrese designed a global quench as follows. Consider a Euclidean cylinder $S^1 \times I$, where $S^1$ is the spatial direction parameterized by $-L/2 < x \leq L/2$ and $I$ is the Euclidean time direction parameterized by $-\alpha < \tau < \alpha$. We consider the thermodynamic limit $L\rightarrow\infty$, where the cylinder becomes an infinite strip. Such a BCFT describes the following pure state at $\tau =0$
\begin{align}\label{eq:quench_initial_state_unitary}
	\rho = e^{-\alpha H }|B\rangle\langle B|e^{-\alpha H}\,,
\end{align}
which is taken to be the initial state of the quench dynamics.
More generally, the strip BCFT realizes a matrix
\begin{align}
    \rho(\tau) = e^{-\tau H }e^{-\alpha H }|B\rangle\langle B|e^{-\alpha H}e^{+\tau H }\,,
\end{align}
at a general $\tau$ slice. Therefore, one can consider the real time evolution under the CFT Hamiltonian by performing an analytic continuation $\tau \rightarrow it$. The energy density evaluated on the infinite strip reads
\begin{align}\label{eq:T_CC}
\langle{T_{\tau\tau}}\rangle_{\rm strip} =-\langle{T_{xx}}\rangle_{\rm strip} =\frac{c}{96}\frac{\pi}{\alpha^2}=\frac{c}{6}\frac{\pi}{(4\alpha)^2}\,,
\end{align}
which is the same as a canonical Gibbs state with inverse temperature
\begin{align}
    \beta = 4\alpha\,.
\end{align}
In chaotic CFT, the state will evolve to a thermodynamic equilibrium with this temperature.
If we instead take the interval $\tau \in [-\alpha,\alpha]$ as our spatial slice, then Eq.\eqref{eq:T_CC} is nothing but the Casimir energy density. In this case, since $|B\rangle$ and $\langle B|$ are described by the same boundary condition, they are connected by the identity operator with $h=0$ at $x\rightarrow \pm \infty$.

Let us then generalize this quench setup to the case with non-unitary boundary conditions.

In a general 2D CFT, a general conformal boundary condition can be represented as a boundary state $|B\rangle$ living in the closed string Hilbert space. It satisfies exactly a half \cite{Ishibashi_1988_boundary_states} of the conformal symmetry as
\begin{align}
	\left(L_n - \bar{L}_{-n}\right) |B\rangle = 0\,,~~~~~n\in\mathbb{N}\,,
\end{align}
where $L_n$ is the Virasoro generator. A general boundary state $|B\rangle$ satisfying this equation can be written as
\begin{align}
	|B\rangle = \sum_j D_{j\mathbb{I}}|j\rangle\rangle\,,
\end{align}
where $|j\rangle\rangle$ is the boundary Ishibashi state associated with the bulk primary labeled by $j$ \cite{Ishibashi_1988_boundary_states}, and the coefficient $D_{j\mathbb{I}}$ is called the disk 1-point function. In Cardy's regime, one further requires the unitarity of the dual open-string sector to be unitary to further constraint the coefficients $D_{j\mathbb{I}}$ \cite{Cardy_1989}, in which $D_{j\mathbb{I}}$'s turn out to be real numbers. Thanks to this feature, the Hermitian conjugate\footnote{Throughout this section, the Hermitian conjugate refers to that on the closed-string Hilbert space. } of $|B\rangle$ reads
\begin{align}
	\left(|B\rangle\right)^{\dagger} = \sum_j D_{j\mathbb{I}}\, \langle\langle j| = \langle B|\,,
\end{align}
and hence \cite{Calabrese_Cardy_2005_global_quench} could use the same boundary condition to describe both the bra and the ket.

In the previous sections, we have constructed physical conformal boundary conditions which do not satisfy the unitary condition in the open-string sector. For these $|B\rangle$, $D_{j\mathbb{I}}$ are in general complex, which admits a conjugate boundary condition
\begin{align}
	|\bar{B}\rangle = \sum_j D_{j\mathbb{I}}^*|j\rangle\rangle\,,
\end{align}
with which
\begin{align}
	\langle \bar{B}| = \left(|B\rangle\,\right)^{\dagger}.
\end{align}
In other words, a non-unitary boundary state and its Hermitian conjugate are described by different boundary conditions.

Now, let us now use such a non-unitary boundary condition to prepare the initial state of the global quench. Then at $\tau=0$, the density matrix reads
\begin{align}\label{eq:quench_initial_state}
	\rho = e^{-\alpha H }|B\rangle\langle \bar{B}|e^{-\alpha H}\,,
\end{align}
This is described by a BCFT defined on the same infinite strip. However, here we impose the boundary condition $B$ at $\tau = -\alpha/2$ and $\bar{B}$ at $\tau = +\alpha/2$.

Since $B$ and $\bar{B}$ are two different boundary conditions, they should be connected by nontrivial bcc operators at $x \rightarrow \pm \infty$. Using $h_{\rm bcc}$ to denote its conformal dimension, the energy density evaluated in this BCFT reads,
\begin{align}\label{eq:T_new}
&\langle{T_{\tau\tau}\Psi(-\infty)\Psi(\infty)}\rangle_{\rm strip} =-\langle{T_{xx}\Psi(-\infty)\Psi(\infty)}\rangle_{\rm strip} \nonumber\\
=&\left(\frac{c}{24}-h_{\rm bcc}\right)\frac{\pi}{4\alpha^2}\,.
\end{align}
Therefore, the effective inverse temperature $\beta$ satisfies
\begin{align}
    \left(\frac{c}{24}-h_{\rm bcc}\right)\frac{\pi}{4\alpha^2} = \frac{c}{6} \frac{\pi}{\beta^2}\,,
\end{align}
and hence
\begin{align}
    \beta =  \frac{4\alpha}{\sqrt{1-24h_{\rm bcc}/c}}\,.
\end{align}
As a result, by using non-unitary boundary conditions, we can have a large class of generalized global quench setups, whose effective temperature deviates from that of the original Cardy-Calabrese one. This is similar to a relation found in the analysis of the entanglement phase transition of pseudo entropy \cite{Kanda_Takayanagi_Wei_2026,pseudoentropy_2020}, which is another context.

Furthermore, since the open-string theory is non-unitary, $h_{\rm bcc}$ can be negative, which implies an enhancement in the effective temperature.
This enhancement of the effective temperature is crucial in understanding a long-standing puzzle. It is known that a conformal boundary state can be regarded as a very special product state where there is no spatial entanglement \cite{MRTW14}, but how special is it?
Consider the canonical Gibbs state with inverse temperature $\beta$, and perform the following decomposition,
\begin{align}\label{eq:METTS}
    \rho_\beta = e^{-\beta H} = \sum_i e^{-\beta H/2 } |P_i\rangle \langle P_i| e^{-\beta H/2 }\,,
\end{align}
where $|P_i\rangle$ is a product state labeled by $i$ and $\sum_i |P_i\rangle \langle P_i| = \mathbb{I}$ is a resolution of the identity with all the product states. Here, the state of the form
\begin{align}
    |\mu_i\rangle = e^{-\alpha H}|P_i\rangle\,,
\end{align}
is called a minimally entangled typical thermal state (METTS) \cite{METTS09}.

From Eq.\eqref{eq:METTS}, it is straightforward to see that a typical METTS with Euclidean time evolution $\alpha$ has the effective temperature $\beta = 2\alpha$. On the other hand, for a unitary boundary state $|B\rangle$, $e^{-\alpha H}|B\rangle$ is also a METTS, but its effective temperature is $\beta = 4\alpha$. This factor of $2$ appearing in the effective inverse temperature implies that $e^{-\alpha H}|B\rangle$ is a very atypical METTS, as argued in Ref.~\cite{METTS2023}.

This itself is not strange, since $|B\rangle$ preserves so many symmetries, which is not preserved by a generic METTS. But is this the whole story? Our analysis above suggests the answer is no. Besides unitary conformal boundary conditions, there also exist physical non-unitary ones. Therefore, even highly symmetric conformal boundary states can have a wide spectrum of effective temperature.

Especially, since a typical METTS has effective inverse temperature $\beta = 2\alpha$, it is plausible to expect the existence of such a non-unitary boundary state. This would imply the existence of a bcc operator with $h_{\rm bcc} = -c/8$. Interestingly, it is recently found that AdS$_3$/CFT$_2$ with dS$_2$ branes admit such non-unitary boundary conditions \cite{HJW26}.
It would be interesting to explore the existence of such bcc operators in more general physical systems.

\section{Discussion and conclusion}

In this work, we study non-unitary conformal interfaces in two-dimensional free CFTs. Our investigation begins with lattice models hosting non-unitary interfaces that are exactly solvable and exactly conformal at the microscopic level. These models are constructed via analytic continuation of known unitary exact conformal interfaces, under which the Hermitian Hamiltonian associated with a unitary conformal interface generally becomes non-Hermitian. In particular, for the free theories studied here, we observe a mode-independent scattering matrix for single-particle wave functions in the lattice models, which provides conclusive evidence for the absence of any scale at the interface that ensures the exact conformality. This gives a concrete physical realization of non-unitary conformal interfaces, avoiding ambiguities associated with non-unitary RG analysis, which remains an open question.

In particular, in the main text we focus on the free fermionic chain with local defects, whose bulk realizes a Dirac fermion CFT at low-energy limit, and the local defects realize exact conformal interface at the lattice level. The lattice model is related to the defect-free Hermitian Hamiltonian via a similarity transformation, which ensures a real spectrum for well-defined low-energy limit and ground state. In particular, under the analytic continuation of the interface parameter, the familiar $SU(2)$ family of unitary conformal interfaces in Dirac fermion CFT becomes $SL(2,\mathbb C)$-parameterized in the non-Hermitian regime. Starting from the exact conformal-interface data, we then construct the corresponding field-theoretic description in both the folded boundary-CFT picture and the unfolded interface-CFT picture. A central lesson from the non-Hermitian lattice Hamiltonian is that the conventional closed-string Hilbert-space construction and open–closed duality must be generalized in a biorthogonal manner. This point becomes explicit when we consider a circle with two symmetric conformal interfaces: after the folding trick, the two interfaces are mapped to incoming and outgoing boundary states in the closed-string channel. We find that consistency with the open-string spectrum is achieved only if these two boundary states are defined independently from the scattering data of the right and left eigenmodes of the non-Hermitian open-string Hamiltonian. Thus, non-Hermiticity does not merely deform the boundary data, but requires a biorthogonal structure that ensures the agreement between the open- and closed-string amplitudes as two representations of the same partition function. Moreover, open–closed duality determines that the boundary entropy takes the same value throughout the entire $SU(2)$ and $SL(2,\mathbb C)$ families of conformal boundary conditions. However, as discussed in Appendix~\ref{app:RG}, from a holographic perspective the standard $g$-theorem~\cite{1991_Affleck_Ludwig, Friedan_2003, Casini_2016_g_theorem} cannot be straightforwardly extended from unitary to non-unitary boundaries. Consequently, the significance of this constant boundary entropy, especially its possible relation to RG monotonicity for non-unitary boundary conditions, remains unclear and requires a separate study.

Our study also reveals several nontrivial features associated with non-unitary conformal interfaces, including the breaking of probability-current conservation and logarithmic entanglement entropy governed by a generally complex-valued effective central charge $c_{\rm eff}$. We find that, when the interface parameter is analytically continued to complex values, the critical logarithmic scaling of the entanglement entropy persists, and the closed expression of $c_{\rm eff}$ previously obtained for unitary conformal interfaces still holds after analytic continuation. This result is established analytically in both the lattice model and the interface CFT using the replica trick, and is further supported by numerical results of lattice simulations. Remarkably, for the free Dirac fermion studied in the main text, there exists an $SU(1,1)$ subfamily of non-unitary interfaces within $SL(2,\mathbb C)$ for which $c_{\rm eff}$ remains real but exhibits unbounded growth as the interface parameter increases. This behavior corresponds to an amplification of single-particle wavefunctions across the interface in the $SU(1,1)$ setting.

Given the existence of such non-unitary conformal boundary conditions, we generalized the global quench setup to the cases where the initial state is prepared with a non-unitary conformal boundary condition and analyzed some universal features. Especially, the incoming and outgoing states are connected by a boundary condition changing operator. We found a formula relating the effective temperature of the global quench and the conformal dimension of the boundary condition changing operator, which can be negative in general and results in a temperature enhancement. A detailed study of these phenomena (quench) in explicit lattice realizations is an interesting direction for future work and will be reported elsewhere.

In addition to the Dirac fermion case presented in the main text, we also study its free scalar analogue in the Appendix~\ref{Appendix:Boson}. The bosonic lattice model is a harmonic-oscillator chain with a local defect, whose structure closely parallels that of the free fermionic chain discussed in the main text. In the unitary case, it realizes an $SO(2)$ family of conformal interfaces. After analytic continuation of the interface parameters, the Hermitian defect Hamiltonian becomes non-Hermitian, and the interface are complexified to an $SO(2,\mathbb{C})$ family. The corresponding field-theoretic description is constructed in close analogy with the Dirac fermion case. For the free boson, we again observe logarithmic entanglement entropy governed by an analytically computable effective central charge, although its asymptotic behavior differs slightly due to the bosonic nature of the theory. Moreover, in Appendix~\ref{app:ising}, we study the free Majorana case, which can be understood as the low-energy theory of an energy defect in the critical Ising model. Although this case does not admit an exact lattice realization of the conformal interface, we analyze the structure of non-unitary conformal interfaces in the free Majorana CFT obtained via an analytic continuation, again corresponding to a generalization from an $SO(2)$ to an $SO(2,\mathbb{C})$ family. The entanglement entropy and effective central charge are derived in a manner similar to the Dirac fermion case, yielding one half of the Dirac-fermion result for $c_{\rm eff}$. This factor of one half is the consequence of that the Majorana theory contains a single real fermionic degree of freedom, whereas the Dirac fermion is complex and consists of two Majorana sectors.

It is worth noting that all non-unitary conformal interfaces considered here are constructed by complexifying the parameters that characterize the corresponding unitary interfaces. The scope of this construction is therefore fundamentally constrained by the symmetry structure of the original unitary interface. In the Dirac fermion case, both the transmission/reflection amplitudes and their phases are tunable, so that complexifying the interface data leads to a general $SL(2,\mathbb C)$ family of non-unitary interfaces. By contrast, in the free boson and free Majorana theories, the allowed unitary interfaces belong only to an $O(2)$ family, with a single parameter controlling the transmission and reflection amplitudes. Consequently, their analytic continuation does not produce a general $SL(2,\mathbb C)$ structure, but only the complexified $O(2,\mathbb C)$ family of interfaces, as discussed in detail in the Appendices. This observation points to a structural limitation of the present complexification approach and raises a natural open question: whether $SL(2,\mathbb C)$ conformal interfaces admit physically relevant lattice realizations in free boson and Majorana/Ising models, possibly through constructions not obtainable by analytic continuation of the known unitary interfaces. More broadly, since this complexification procedure requires the underlying unitary interface to possess continuously tunable parameters, it remains an important question how to construct non-unitary boundaries and interfaces in minimal models, where continuous families of unitary conformal boundaries and interfaces are generally absent.

Several questions remain concerning the microscopic realization of the lattice--field-theory correspondence for the non-unitary interfaces studied in this work. One important direction is to understand how the residual conservation laws of non-unitary interfaces are encoded at the lattice level. In the Dirac fermion example, the $SU(1,1)$ subfamily does not conserve the ordinary probability current, but instead preserves a pseudo-current associated with the indefinite structure of $SU(1,1)$. Clarifying the lattice counterpart of this pseudo-current, as well as its precise matching to the conserved quantity in the continuum interface CFT, would sharpen the physical interpretation of non-unitary interfaces with non-compact group structure.

Another worth exploring direction is to construct the non-unitary boundary data directly from microscopic Hamiltonians. In particular, one may ask whether the biorthogonal boundary states appearing in the closed-string channel can be realized as the left and right ground-state data of suitable non-Hermitian massive Hamiltonians. Such a construction would provide a concrete open-channel realization of the corresponding non-unitary boundary conditions, and would offer a controlled framework for studying both equilibrium and non-equilibrium phenomena associated with them.

Moreover, it would be valuable to extend the entanglement analysis of Sec.~\ref{Sec:EE} beyond ground states. There, the entanglement entropy across a conformal interface is obtained by replacing the central charge $c$ in the universal logarithmic term with the effective central charge $c_{\mathrm{eff}}$ associated with the interface. Applying the methods developed in Ref.~\cite{Eisler_2012_evo_defect,2018_Wen,2022_Eisler}, one can further show that the same $c_{\mathrm{eff}}$ governs nonequilibrium settings, including global and local quantum quenches. Furthermore, as will be shown in Ref.~\cite{2026_DrivenCFT}, $c_{\mathrm{eff}}$ also controls the entanglement entropy dynamics in time-dependent driven conformal field theories. It would be interesting to establish these extensions systematically from both lattice and field-theory perspectives.

It would also be interesting to explore the relation between the non-unitary boundary states and topological phases of matter. For the Dirac fermion CFT, the family of unitary conformal boundary states parametrized by $SU(2)\simeq S^3$ was previously shown to support a nontrivial higher Berry phase characterized by a quantized Dixmier-Douady invariant \cite{2025_Choi_Ryu,2025_Wen}. The present work naturally enlarges the parameter space of exactly solvable conformal interfaces and boundary conditions from $SU(2)$ to $SL(2,\mathbb C)$. This enlarged parameter manifold may support structures beyond the higher Berry curvature/connection in the unitary case.

Last but not least, our findings suggest a possible way to classify critical points in open quantum systems. While classification of critical behaviors in closed quantum many-body systems is well-understood with the tool of unitary CFTs, that in open quantum many-body systems remains unsolved. In particular, even critical behaviors in 1D non-Hermitian quantum many-body systems, which is a small subclass, is not well-understood. Our analysis suggests there exists a even simpler case: 0D non-Hermitian defects in 1D Hermitian theories. In this case, one may leverage the power of the unitarity of the parent CFT to accomplish a classification.
Understanding this simpler setting may therefore provide valuable insights into the more general problem of classifying criticality in genuinely non-Hermitian one-dimensional systems.

\medskip

\textit{Note added}:
During the final preparation of this manuscript, Ref.~\cite{2026_Takayanagi} appeared on arXiv. We also became aware of another related work~\cite{2026_Yuya}, which is coordinated and will appear on arXiv on the same day.

~\par
\begin{acknowledgments}
This work is supported by a startup at Georgia Institute of Technology. ZW is supported by the Society of Fellows at Harvard University. Q.T. especially thanks Jiannan Hua for sharing the note on unitary conformal interface in free fermionic chain. We thank Hosho Katsura, Kohei Kawabata, Jiaxin Qiao, Kazuaki Takasan, Tadashi Takayanagi
and Carlos Sa de Melo for interesting discussions.
We also thank the authors of Ref.~\cite{2026_Yuya} for coordinating submissions of our papers to arXiv.
\end{acknowledgments}


\appendix

\section{Exactly solvable non-unitary conformal interfaces in free boson/scalar theory}
\label{Appendix:Boson}

In the main text, we constructed exactly solvable non-unitary conformal interfaces for a free Dirac fermion. In this appendix, we show that a closely related construction also works for a free scalar theory, realized on the lattice by a harmonic oscillator chain.

\subsection{Exact conformal interface of a harmonic oscillator chain}

Consider a chain of $2L$ coupled harmonic oscillators~\cite{Eisler_2012_defect_boson},
\begin{equation}
	\begin{aligned}
		\hat{\mathbf H}_{\rm boson}
		&=
		\sum_{n=1}^{2L}
		\left(
		-\frac{1}{2m_n}\frac{\partial^2}{\partial x_n^2}
		+
		\frac{1}{2}m_n\Omega_0^2x_n^2
		\right)
		\\ & \quad +
		\frac{1}{2}
		\sum_{n=1}^{2L-1}
		K_n(x_n-x_{n+1})^2 .
	\end{aligned}
\end{equation}
Here $\Omega_0$ is the bare oscillator frequency, $m_n$ is the site-dependent mass, and $K_n$ is the spring constant on the bond $(n,n+1)$. We choose
\begin{equation}
	m_n=
	\begin{cases}
		K_{\rm L}=e^\varphi, & 1\le n\le L,\\
		K_{\rm R}=e^{-\varphi}, & L+1\le n\le 2L,
	\end{cases}
\end{equation}
with
\begin{equation}
K_n=
\begin{cases}
	K_{\rm L}, & n<L,\\
	K_c, & n=L,\\
	K_{\rm R}, & n>L,
\end{cases}
\end{equation}
and
\begin{equation}
	K_c=\frac{2K_{\rm L}K_{\rm R}}{K_{\rm L}+K_{\rm R}}=\sech\varphi .
\end{equation}
After rescaling the coordinates as $u_n=\sqrt{m_n}x_n$, the Hamiltonian becomes
\begin{equation}
	\begin{aligned}
		&\hat{\mathbf H}_{\rm boson, rescale}
		=
		\sum_{n=1}^{2L}
		\left(
		-\frac{1}{2}\frac{\partial^2}{\partial u_n^2}
		+
		\frac{1}{2}\Omega_0^2u_n^2
		\right)
		\\ & +
		\frac{1}{2}\sum_{n=1}^{L-1}(u_n-u_{n+1})^2
		+
		\frac{1}{2}\sum_{n=L+1}^{2L-1}(u_n-u_{n+1})^2
		\\
		&
		+
		\frac{1}{2}
		\left(
		\frac{K_c}{K_{\rm L}}u_L^2
		+
		\frac{K_c}{K_{\rm R}}u_{L+1}^2
		-
		2\frac{K_c}{\sqrt{K_{\rm L}K_{\rm R}}}u_Lu_{L+1}
		\right).
	\end{aligned}
\end{equation}
Thus the bulk is homogeneous after rescaling, while the inhomogeneity is localized at the central bond.

It is useful to write the Hamiltonian in matrix form,
\begin{equation}
	\hat{\mathbf H}_{\rm boson}
	=
	\frac{1}{2}\mathbf P^T M^{-1}\mathbf P
	+
	\frac{1}{2}\mathbf X^T\Phi\mathbf X,
\end{equation}
where
\begin{equation}
	\mathbf P=
	\left(
	-i\frac{\partial}{\partial x_1},\ldots,
	-i\frac{\partial}{\partial x_{2L}}
	\right),
	\quad
	\mathbf X=(x_1,\ldots,x_{2L})^T .
\end{equation}
The nonzero elements of $M$ and $\Phi$ are
\begin{equation}
\begin{aligned}
	M_{n,n}&=m_n,\\
	\Phi_{n,n+1}&=\Phi_{n+1,n}=-K_n,\\
	\Phi_{n,n}
	&=
	\begin{cases}
		m_n\Omega_0^2+K_n, & n=1,\\
		m_n\Omega_0^2+K_{n-1}, & n=2L,\\
		m_n\Omega_0^2+K_{n-1}+K_n, & 2\le n\le 2L-1.
	\end{cases}
\end{aligned}
\end{equation}
The relevant dynamical matrix is
\begin{equation}
	D=M^{-\frac12}\Phi M^{-\frac12},
\end{equation}
which governs the equations of motion for the rescaled variables
\begin{equation}
	\frac{\partial^2}{\partial t^2}\mathbf U=-D\mathbf U,
	\qquad
	\mathbf U=M^{\frac12}\mathbf X=(u_1,\ldots,u_{2L})^T .
\end{equation}

In the homogeneous case, the eigenfrequencies are
\begin{equation}
	\Omega_k^2
	=
	\Omega_0^2
	+
	2\left(
	1-\cos\theta_k
	\right),
	\quad \frac{k\pi}{2L} ,
	\quad
	k=0,\ldots,2L-1,
\end{equation}
with eigenfunctions
\begin{equation}\label{eq:boson_homogenous_eigenfunc}
	\phi_k^0(n)
	=
	\begin{cases}
		\sqrt{\frac{1}{L}}\,
		\cos[(n-\frac12)\theta_k],
		& k\neq0,\\[0.5ex]
		\sqrt{\frac{1}{2L}},
		& k=0.
	\end{cases}
\end{equation}
The overlap matrix restricted to the left half of the chain is~\cite{Eisler_2009_RDM}
\begin{equation}
	A^0_{kq}
	=
	\sum_{n=1}^{L}\phi_k^0(n)\phi_q^0(n).
\end{equation}

For the inhomogeneous chain, which converts to the defected Hamiltonian upon a rescaling, the eigenfrequencies remain unchanged, while the eigenfunctions are rescaled on the two sides of the interface:
\begin{equation}\label{eq:boson_defect_eigenfunc}
	\widetilde\phi_k(n)
	=
	\begin{cases}
		\alpha_k\phi_k^0(n), & 1\le n\le L,\\
		\beta_k\phi_k^0(n), & L+1\le n\le 2L,
	\end{cases}
\end{equation}
with
\begin{equation}
	\alpha_k^2=1+(-1)^k\tanh\varphi,
	\qquad
	\beta_k^2=1-(-1)^k\tanh\varphi .
\end{equation}
The branch is chosen such that
\begin{equation}
	\alpha_k\beta_k=\sech\varphi .
\end{equation}
The corresponding left-half overlap matrix is
\begin{equation}
	\widetilde A_{kq}
	=
	\alpha_k\alpha_q A^0_{kq}.
\end{equation}

The fact that the interface only modifies the relative amplitudes on the two sides, without changing the frequency quantization, already suggests that the interface scattering is momentum independent. To make this explicit, for $k\neq0$ we define the lattice incoming and outgoing modes near the interface as
\begin{equation}\label{eq:boson_lattice_mode}
	\begin{aligned}
		b_k^{{\rm I},{\rm in}}
		&=
		-i e^{\frac{3i}{2}\theta_k}
		\left[
		\phi_k(L)-e^{-i\theta_k}\phi_k(L-1)
		\right],
		\\
		b_k^{{\rm I},{\rm out}}
		&=
		i e^{-\frac{3i}{2}\theta_k}
		\left[
		\phi_k(L)-e^{i\theta_k}\phi_k(L-1)
		\right],
		\\
		b_k^{{\rm II},{\rm in}}
		&=
		-i e^{\frac{3i}{2}\theta_k}
		\left[
		\phi_k(L+1)-e^{-i\theta_k}\phi_k(L+2)
		\right],
		\\
		b_k^{{\rm II},{\rm out}}
		&=
		i e^{-\frac{3i}{2}\theta_k}
		\left[
		\phi_k(L+1)-e^{i\theta_k}\phi_k(L+2)
		\right],
	\end{aligned}
\end{equation}
Substituting these modes into the interface equation gives the momentum-independent scattering matrix
\begin{equation}
	S_{\rm boson}(k)=S_{\rm boson}
	=
	\begin{pmatrix}
		\tanh\varphi & \sech\varphi\\
		\sech\varphi & -\tanh\varphi
	\end{pmatrix}
	\in O(2).
\end{equation}
Equivalently,
\begin{equation}
	\hat S_{\rm boson}
	=
	iS_{\rm boson}
	\in SO(2).
\end{equation}
The free boson conformal interface is therefore described by an $O(2)$ scattering matrix. The present lattice construction realizes the branch with $\det S_{\rm boson}=-1$.

As in the Dirac fermion case, we can analytically continue the interface parameter to complex values, $\varphi\in\mathbb C$. The eigenvalue equations for the right eigenmodes keep the same form, and hence
\begin{equation}
	\widetilde\phi^{\mathbf r}_k(n)
	=
	\begin{cases}
		\alpha_k\phi_k^0(n), & 1\le n\le L,\\
		\beta_k\phi_k^0(n), & L+1\le n\le 2L,
	\end{cases}
\end{equation}
with the same branch condition $\alpha_k\beta_k=\sech\varphi$. The left eigenmodes are obtained from the Hermitian conjugate problem, which amounts here to complex conjugating the interface parameter:
\begin{equation}
	\widetilde\phi^{\mathbf l}_k(n)
	=
	\begin{cases}
		\alpha_k^*\phi_k^0(n), & 1\le n\le L,\\
		\beta_k^*\phi_k^0(n), & L+1\le n\le 2L.
	\end{cases}
\end{equation}
Accordingly, the right-eigenmode scattering matrix is
\begin{equation}\label{eq:nonunitary_boson_scattering_right}
	S^{\mathbf r}_{\rm boson}
	=
	S_{\rm boson}
	=
	\begin{pmatrix}
		\tanh\varphi & \sech\varphi\\
		\sech\varphi & -\tanh\varphi
	\end{pmatrix}
	\in O(2,\mathbb C),
\end{equation}
whereas the left-eigenmode scattering matrix is
\begin{equation}\label{eq:nonunitary_boson_scattering_left}
	S^{\mathbf l}_{\rm boson}
	=
	\bar S_{\rm boson}
	=
	\begin{pmatrix}
		\tanh\varphi^* & \sech\varphi^*\\
		\sech\varphi^* & -\tanh\varphi^*
	\end{pmatrix}
	\in O(2,\mathbb C).
\end{equation}
Thus
\begin{equation}
	\hat S_{\rm boson}^{\,{\rm non\text{-}unitary}}
	=
	iS_{\rm boson}
	\in SO(2,\mathbb C).
\end{equation}
Unlike the free Dirac fermion case, the analytic continuation considered here keeps the bosonic interface within the complex orthogonal group. It therefore does not generate a general $SL(2,\mathbb C)$ family of conformal interfaces.

One can also compute the half-cut entanglement entropy for the subsystem $A=[1,L]$. For the unitary interface, the relevant bosonic correlation matrix $C=2\mathbf{X}2\mathbf{P}$ satisfies
\begin{equation}\label{eq:boson_correlation_matrix}
\begin{aligned}
	\widetilde C_A
	& =
	(1-\tanh^2\varphi) \, C^0_A
	+
	\tanh^2\varphi\,\Id
	\\ & =
	s^2 \, C^0_A+(1-s^2)\Id \,,
\end{aligned}
\end{equation}
where $s=\sech\varphi$.
For a bosonic Gaussian state, if $\xi_l^2$ denotes an eigenvalue of $C_A$, then
\begin{equation}
	\xi_l^2
	=
	\coth^2\frac{\epsilon_l}{2},
\end{equation}
where $\epsilon_l$ is the corresponding single-particle entanglement energy. Equation \eqref{eq:boson_correlation_matrix} therefore implies
\begin{equation}\label{eq:relation_ES_boson_defect}
	\sinh\frac{\widetilde\epsilon_l}{2}
	=
	\frac{1}{s}
	\sinh\frac{\epsilon^0_l}{2}.
\end{equation}
Under analytic continuation, this becomes
\begin{equation}
	\sinh\frac{\widetilde\epsilon_l}{2}
	=
	\frac{1}{\sqrt{s^2}}
	\sinh\frac{\epsilon^0_l}{2}.
\end{equation}

The entanglement entropy takes the form
\begin{equation}
	S_A
	=
	\frac{I^{\rm boson}(\mathcal S)}{2\pi^2}
	\log\frac{L}{\epsilon}
	+
	O(1),
	\qquad
	\mathcal S=\sqrt{s^2},
\end{equation}
where
\begin{equation}
	I^{\rm boson}(\mathcal S)
	=
	-\int_0^\infty d\epsilon^0\,
	\log\!\left(1-e^{-\widetilde\epsilon}\right)
	+
	\int_0^\infty d\epsilon^0\,
	\frac{\widetilde\epsilon}{e^{\widetilde\epsilon}-1}.
\end{equation}
This gives
\begin{equation}
	\begin{aligned}
		c_{\rm eff}^{\rm boson}
		&=
		\frac{3}{2}\mathcal S
		+
		\frac{3}{\pi^2}
		\sum_{\kappa=\pm1}
		\Big[
		(1+\kappa\mathcal S)\,
		{\rm Li}_2(-\kappa\mathcal S)
		\\
		&\hspace{3.0cm}
		+
		(1+\kappa\mathcal S)\,
		\log(1+\kappa\mathcal S)\log\mathcal S
		\Big].
	\end{aligned}
\end{equation}
Comparing with the Dirac result in \eqref{eq:effective_central_charge}, one obtains \cite{Eisler_2012_defect_boson}
\begin{equation}\label{eq:boson_lattice_effective_central_charge}
	c_{\rm eff}^{\rm boson}(\mathcal S)
	=
	\frac{3}{2}\mathcal S
	-
	\frac{1}{2}c_{\rm eff}^{\rm Dirac}(\mathcal S).
\end{equation}
For real positive $\mathcal S$, the large-$\mathcal S$ behavior is therefore
\begin{equation}
	c_{\rm eff}^{\rm boson}(\mathcal S)
	=
	O\left( (\log\mathcal S)^2 \right),
	\qquad
	\mathcal S\to\infty,
\end{equation}
which is qualitatively different from the linear growth found for the Dirac fermion.

\subsection{Boundary and interface CFT description of the bosonic interface}

Analogously to the free Dirac fermion discussed in the main text, the exact bosonic lattice interface admits a boundary/interface CFT description. Since the formalism is very similar, we only present the main results. For simplicity, we consider a non-compact boson that directly corresponds to the harmonic oscillator chain.

Consider the unfolded theory on $x\in[-L,L]$, with an interface at $x=0$. Folding along the interface maps the problem to a boundary condition of the product theory
$\mathrm{CFT}^{\rm I}\otimes\overline{\mathrm{CFT}}{}^{\rm II}$ on $x\in[-L,0]$. In the closed-string channel, the right-eigenmode gluing condition is
\begin{equation}
	\partial\Phi_+(w')
	=
	S_{\rm boson}\,\partial\Phi_-(w'),
\end{equation}
where $\Phi_\pm=(\Phi_\pm^{\rm I},\Phi_\pm^{\rm II})^T$ are the chiral and anti-chiral scalar fields of the folded theory. Equivalently,
\begin{equation}
	\lim_{\Im w'\to0}
	\llangle a|
	\left[
	\partial\Phi_+(w')
	-
	S_{\rm boson}\partial\Phi_-(w')
	\right]
	|b\rrangle
	=
	0.
\end{equation}
The incoming boundary state is
\begin{equation}
	|b\rrangle
	=
	g_b
	\exp\!\left[
	\sum_{n=1}^{\infty}
	\frac{1}{n}
	a_{-n}^T S_{\rm boson}\bar a_{-n}
	\right]
	|\mathcal{B}_{\rm zero}\rangle,
\end{equation}
where $a_n=(a_n^{\rm I},a_n^{\rm II})^T$ and
$\bar a_n=(\bar a_n^{\rm I},\bar a_n^{\rm II})^T$ are oscillator modes of the chiral and anti-chiral bosons, and $|\mathcal{B}_{\rm zero}[S_{\rm boson}]\rrangle$ contains the zero-mode part of the bosonic boundary state. For the non-compact boson studied here, the logarithmic coefficient of the entanglement entropy is determined by the oscillator sector, therefore we are not going to discuss the detailed zero-mode contribution below. The outgoing state is
\begin{equation}
	\llangle b|
	=
	g_b^*
	\langle\mathcal{B}_{\rm zero}|
	\exp\!\left[
	\sum_{n=1}^{\infty}
	\frac{1}{n}
	\bar a_n^T S_{\rm boson}^\dagger a_n
	\right].
\end{equation}
For the non-unitary interface, the left-eigenmode boundary states are obtained by replacing $S_{\rm boson}$ with $\bar S_{\rm boson}$:
\begin{equation}
	|\bar b\rrangle=|b\rrangle(S_{\rm boson}\to\bar S_{\rm boson}),
	\
	\llangle\bar b|
	=
	\llangle b|(S_{\rm boson}\to\bar S_{\rm boson}).
\end{equation}
The matrices $S_{\rm boson}$ and $\bar S_{\rm boson}$ are precisely the right- and left-eigenmode scattering matrices in
\eqref{eq:nonunitary_boson_scattering_right} and
\eqref{eq:nonunitary_boson_scattering_left}.

Unfolding these boundary states gives the corresponding interface operators. For $\mathcal I_{12}^{\rm boson}:\mathcal H_{\rm II}^{\rm boson}\to\mathcal H_{\rm I}^{\rm boson}$, it is given by
\be
\mathcal{I}_{12}^{\rm boson} = \left[ \prod_{r=1}^{\infty} \mathcal{I}_{12}^{\rm boson}(r) \right] \mathcal{I}_{12}^{\,\rm zero} \,
\ee
with the oscillating part
\be
\mathcal{I}_{12}^{\rm boson}(r) = \exp \left[ \frac{1}{r} \begin{pmatrix}
    a_{-r}^{\rm I} & \bar{a}_r^{\rm II}
\end{pmatrix} D_{\rm boson}
\begin{pmatrix}
    \bar{a}_{-r}^{\rm I} \\ a_r^{\rm II}
\end{pmatrix} \right]
\ee
where
\be
D_{\rm boson} = \begin{pmatrix}
    (S_{\rm boson})_{11} & - (S_{\rm boson})_{12} \\ - (S_{\rm boson})_{21} & (S_{\rm boson})_{22}
\end{pmatrix}
\ee
and $\mathcal{I}_{12}^{\,\rm zero}$ denotes the map from the ground state of CFT$^{\rm II}$ with zero-mode contribution to the one of CFT$^{\rm I}$.

The opposite-orientation interface operator $\mathcal I_{21}^{\rm boson}=:\mathcal H_{\rm I}^{\rm boson}\to\mathcal H_{\rm II}^{\rm boson}$ is given by
\be
\mathcal I_{21}^{\rm boson}=\mathcal I_{12}^{\rm boson}\Big|_{{\rm I} \leftrightarrow {\rm II}, S_{\rm boson} \leftrightarrow S_{\rm boson}^{-1}}
\ee
It is worth noting that, in the constructed non-unitary bosonic interface, one still has
\be
(S^{\rm non-unitary}_{\rm boson})^{-1} = (S^{\rm non-unitary}_{\rm boson})^T
\ee
as the unitary case. This is because that, after the analytic continuation, the non-unitary scattering matrix is in $O(2,\mathbb C)$, i.e. still orthogonal.

Based on the interface operator obtained above, one can follow a similar replica calculation as the Dirac-fermion case discussed in the main text for evaluating EE. Here, for the free boson, it would be useful to consider a linearization by letting
\be
\vec{A}_r = \frac{1}{r} \begin{pmatrix}
    a_{-r}^{\rm I} \\ \bar{a}_{r}^{\rm II}
\end{pmatrix} \,, \qquad
\vec{B}_r = D_{\rm boson} \begin{pmatrix}
    \bar{a}_{-r}^{\rm I} \\ a_r^{\rm II}
\end{pmatrix}
\ee
and using the identity
\be
e^{\vec{A}\cdot\vec{B}} = \int \frac{d^2 \vec{z}}{\pi^2} e^{-\vec{z}\cdot\vec{\bar z} - \vec{z}\cdot\vec{A} - \vec{z}\cdot\vec{B}}
\ee
which is valid when all components of $\vec{A}=(A^1, A^2)^T$ and $\vec{B}=(B^1, B^2)^T$ commute with each other.
Moreover, for free boson one has
\be
e^{a_r^j} q^{\hat{L}_0^j} = q^{\hat{L}_0^j} e^{r \, a_r^j}
\ee
where $\hat{L}_0^j$ denote the zero-th (chiral) Virasoro generator for CFT$^j$, $j={\rm I}, {\rm II}$. This is different from the fermion case, where switching the fermionic operator induces a sign change. In fact, this sign difference is the fundamental reason why EE in free boson behaves differently than free fermion.

After some straightforward calculation analog to the free fermion case, from \eqref{eq:trace_replica_partition_func} one arrives
\be
\mathcal{Z}^{(n)}_{\rm boson} = \mathcal{Z}^{(n)}_{{\rm boson}, {\rm osc}} \mathcal{Z}^{(n)}_{{\rm boson}, {\rm zero}}
\ee
where $n$ is the replica index, and the subscripts ``${\rm osc}$'' and ``${\rm zero}$'' denote the contributions from oscillating and zero modes, respectively. As we have pointed out, for non-compact boson studied here, the zero-mode contribution does not play a role in the leading logarithmic scaling of EE. For the oscillating part, one has
\be
\mathcal{Z}^{(n)}_{{\rm boson}, {\rm osc}} = \prod_{r=1}^{\infty} \prod_{k=1}^{n} \left[ 1 - 2 \cos(2\nu_k) \, \hat{q}^{2r} + \hat{q}^{4r} \right]^{-1}
\ee
where
\be
\nu_k = \arcsin \left[ \mathcal{S} \sin \frac{\pi k}{n} \right] \,,
\quad \mathcal{S} = \sqrt{s^2}\,, \quad s = \sin\varphi
\ee
and the modular nome is defined as
\be
\hat q = \exp(i\pi\tau) = \exp(-\frac{2\pi^2}{W}) \,, \quad W = \log \frac{L}{\epsilon}
\ee
The normalized partition function is then
\be
\begin{aligned}
Z^{(n)} & = \prod_{r=1}^{\infty} \prod_{k=1}^{n} \frac{ (1 - \hat{q}^{2r})^2 }{ 1 - 2 \cos(2\nu_k) \, \hat{q}^{2r} + \hat{q}^{4r} }
\\ & = \prod_{k=1}^{n} \frac{\sin(\nu_k) \,\theta'_1(0|\tau)}{\theta_1(\nu_k|\tau)}
\end{aligned}
\ee
where $\theta'_1(0|\tau) = 2\eta^3(\tau)$. Its asymptotic behavior at large $W$ is given by
\be
\log Z_A^{(n)} \sim - \frac{W}{2\pi^2} \, g(n)
\ee
with
\be
g(n) =
\sum_{k=1}^{n}\left( \pi \nu_k - \nu_k^2 \right) \,.
\ee
It leads to
\be
\begin{aligned}
S = - \partial_n \log Z_A^{(n)}\Big|_{n=1} & \sim \frac{W}{2\pi^2} \frac{\partial g(n)}{\partial n} \Bigg|_{n=1}
\\ & \quad = \frac{c_{\rm eff}^{\rm boson}}{6} \log \frac{L}{\epsilon} + \mathcal{O}(1)
\end{aligned}
\ee
By considering
\be
\begin{aligned}
\partial_n g(n) & =\pi \, \partial_n \left( \sum_{k=1}^{n} \nu_k \right) - \partial_n \left( \sum_{k=1}^{n} \nu_k^2 \right)
\\ & = \pi \, \partial_n \left( \sum_{k=1}^{n} \nu_k \right) - \frac{\partial f(n)}{\partial n}
\end{aligned}
\ee
and
\be
\partial_n \left( \sum_{k=1}^{n} \nu_k \right) \Bigg|_{n=1} = \frac{\pi \mathcal{S}}{2}
\ee
Here the replica limit of $\partial_n f(n)\big|_{n\to1}$ is evaluated in \eqref{eq:replica_limit_integral} for free Dirac fermion.
One has \cite{2008_Sakai}
\begin{equation}
	c_{\rm eff}^{\rm boson}
	=
	\frac{3\mathcal S}{2}
	-
	\frac{3}{\pi^2}
	\int_0^\infty
	\frac{
		4u\,du
	}{
		\exp\!\left[
		2\,\operatorname{arcsinh}
		\left(
		\frac{\sinh u}{\mathcal S}
		\right)
		\right]-1
	},
\end{equation}
or equivalently
\begin{equation}
	c_{\rm eff}^{\rm boson}
	=
	\frac{3\mathcal S}{2}
	-
	\frac{1}{2}c_{\rm eff}^{\rm Dirac}(\mathcal S) ,
\end{equation}
where $c_{\rm eff}^{\rm Dirac}$ is the Dirac-fermion effective central charge in \eqref{eq:effective_central_charge}, and $\mathcal{S}$ plays the same role as the interface operator $\Lambda$ in the expression. This result is identical with the lattice calculation in \eqref{eq:boson_lattice_effective_central_charge}.

\section{Non-unitary conformal interface in the Ising model}\label{app:ising}

We have discussed exact non-unitary conformal interfaces in both the free Dirac fermion and the free boson. In this appendix, we also briefly discuss the corresponding construction for the Ising model. Unlike the previous two examples, the lattice defect considered here is not exactly conformal at finite size. Instead, its low-energy limit gives the conformal interface.

Consider a transverse-field Ising chain with inhomogeneous couplings~\cite{1996_Oshikawa,1997_Oshikawa},
\begin{equation}
	\hat{\mathbf H}_{\rm Ising}
	=
	-\sum_{i=1}^{2L-1}
	J_i\sigma_i^x\sigma_{i+1}^x
	-
	\sum_{i=1}^{2L}
	g_i\sigma_i^z.
\end{equation}
After a Jordan--Wigner transformation, it becomes a Kitaev chain,
\begin{equation}
	\begin{aligned}
		\hat{\mathbf H}_{\rm Kitaev}
		&=
		-\sum_{i=1}^{2L-1}
		J_i
		\left(
		c_i^\dagger c_{i+1}
		+
		{\rm h.c.}
		\right)
		\\
		&\quad
		-
		\sum_{i=1}^{2L-1}
		J_i
		\left(
		c_i^\dagger c_{i+1}^\dagger
		+
		{\rm h.c.}
		\right)
		-
		\sum_{i=1}^{2L}
		g_i(2n_i-1).
	\end{aligned}
\end{equation}
We focus on the critical chain with a bond defect at the center,
\begin{equation}
	g_i=1,
	\qquad
	J_i=
	\begin{cases}
		t, & i=L,\\
		1, & i\neq L.
	\end{cases}
\end{equation}
This lattice defect corresponds in the continuum to an energy-defect line in the two-dimensional Ising model.

In the fermion representation, at the low-energy limit the interface becomes conformal and is characterized by
\begin{equation}
	\begin{aligned}
		S_{\rm Ising}
		&=
		\frac{1}{t^2+1}
		\begin{pmatrix}
			t^2-1 & 2t\\
			2t & -t^2+1
		\end{pmatrix}
		\\
		&=
		\begin{pmatrix}
			-\cos\theta & \sin\theta\\
			\sin\theta & \cos\theta
		\end{pmatrix}
		\in O(2),
	\end{aligned}
\end{equation}
where
\begin{equation}
	\theta=2\arctan t.
\end{equation}
A non-unitary continuation is obtained by allowing $t$, or equivalently $\theta$, to become complex. The right- and left-eigenmode scattering matrices are then
\begin{equation}
	S_{\rm Ising}^{\mathbf r}
	=
	S_{\rm Ising}
	=
	\begin{pmatrix}
		-\cos\theta & \sin\theta\\
		\sin\theta & \cos\theta
	\end{pmatrix}
	\in O(2,\mathbb C),
\end{equation}
and
\begin{equation}
	S_{\rm Ising}^{\mathbf l}
	=
	\bar S_{\rm Ising}
	=
	\begin{pmatrix}
		-\cos\theta^* & \sin\theta^*\\
		\sin\theta^* & \cos\theta^*
	\end{pmatrix}
	\in O(2,\mathbb C).
\end{equation}

The folded boundary states in the NS sector are
\begin{equation}
	\begin{aligned}
		|b\rrangle
		&=
		\exp\!\left[
		-i\sum_{n>0}
		\psi_{-n}^T
		S_{\rm Ising}
		\bar\psi_{-n}
		\right]
		|0\rangle,
		\\
		\llangle b|
		&=
		\langle0|
		\exp\!\left[
		-i\sum_{n>0}
		\bar\psi_n^T
		S_{\rm Ising}^\dagger
		\psi_n
		\right],
	\end{aligned}
\end{equation}
where $\psi_n=(\psi_n^{\rm I},\psi_n^{\rm II})^T$ and
$\bar\psi_n=(\bar\psi_n^{\rm I},\bar\psi_n^{\rm II})^T$ are Majorana modes, with $n\in\mathbb N-\frac12$ as we focus on the NS sector. Nevertheless, the zero-mode contribution in R sector is straightforward to be integrated. The left-eigenmode boundary states are obtained by replacing $S_{\rm Ising}$ with $\bar S_{\rm Ising}$:
\begin{equation}
	|\bar b\rrangle=|b\rrangle(S_{\rm Ising}\to\bar S_{\rm Ising}),
	\
	\llangle\bar b|
	=
	\llangle b|(S_{\rm Ising}\to\bar S_{\rm Ising}).
\end{equation}

Unfolding the boundary states gives the interface operator $\mathcal{I}_{12}^{\rm Majorana} : \mathcal{H}_{\rm II}^{\rm Majorana} \to \mathcal{H}_{\rm I}^{\rm Majorana}$ as
\be
\mathcal{I}_{12}^{{\rm Majorana}} = \left[
	\prod_{r\in\mathbb N-\frac12}
	\mathcal I_{12}^{{\rm Majorana}}(r)
	\right]
	P_{12}^{{\rm Majorana}, {\rm NS}} \,,
\ee
where the map from the ground states of CFT$^{\rm II}$ to CFT$^{\rm I}$ is given by
\begin{equation}
	P_{12}^{{\rm Majorana}, {\rm NS}}
	=
	|0,{\rm NS}\rangle_{\rm I}\,
	{}_{\rm II}\langle0,{\rm NS}|
\end{equation}
Here we focus on the NS sector, while the R sector can be generalized with including the zero-mode contribution. In particular, here we have
\be
I_{12}^{{\rm Majorana}}(r)=I_{12}^{{\rm Dirac}, (+)}
\ee
that discussed in \eqref{eq:I12_matrix} for Dirac fermion with replacing the $(+)$-sector component of the Dirac fermion to be the real Majorana.
The replica calculation is also parallel to the free Dirac case. The only difference is that the Majorana theory has a single real component rather than a complex Dirac fermion, so the replica partition function is~\cite{Brehm2015_EE_interfece_ising}
\begin{equation}
	\begin{aligned}
		Z_{{\rm Ising},A}^{(n)}
		&=
		\prod_{k=1}^{n}
		\prod_{r\in\mathbb N-\frac12}
		\left[
		\frac{
			1+2\cos(2\nu_k)\widehat q^{\,r}
			+\widehat q^{\,2r}
		}{
			(1+\widehat q^{\,r})^2
		}
		\right]
		\\
		&=
		\prod_{k=1}^{n}
		\frac{
			\theta_3(\nu_k|\tau)
		}{
			\theta_3(0|\tau)
		}.
	\end{aligned}
\end{equation}
with
\be
\hat q = \exp(i\pi\tau) = \exp(-\frac{2\pi^2}{W}) \,, \qquad W = \log \frac{L}{\epsilon}
\ee
and
\be
\nu_k = \pm \arcsin \left[ \mathcal{S} \sin \frac{\pi k}{n} \right] \,,
\ee
where
\begin{equation}
	\mathcal S=\sqrt{s^2},
	\qquad
	s=\sin\theta=\sin(2\arctan t),
\end{equation}

Here
\begin{equation}
	\left[ Z_{{\rm Ising}, A}^{(n)} \right]^2 = Z_{{\rm Dirac}, A}^{(n)}
\end{equation}
for reproducing the Dirac-fermion result in \eqref{eq:replica_partition_theta}.
Consequently,
\begin{equation}
	c_{\rm eff}^{\rm Ising}(\mathcal S)
	=
	\frac{1}{2}
	c_{\rm eff}^{\rm Dirac}(\mathcal S),
\end{equation}
where $\mathcal{S}$ plays the same role as $\Lambda=\sqrt{\lambda^2}$ in
\eqref{eq:effective_central_charge}.

\section{Breakdown of the the RG monotonicity through a holographic lens}\label{app:RG}

It is known that the $g$-function as well as the boundary entropy monotonically decrease under the usual boundary RG flow. This is known as the $g$-theorem~\cite{1991_Affleck_Ludwig, Friedan_2003, Casini_2016_g_theorem}. Given the existence of non-unitary boundary conditions, one may wonder if there is any form of monotonicity under RG for their boundary entropy. In this appendix, we argue that, at least for $|g|$, its value can both increase and decrease under natural RG flows, by looking at a simple toy model in AdS/CFT. 

Before going into the non-unitary boundary conditions, let us briefly recast what the counterpart of the $g$-theorem was on the AdS side. A simple gravity dual of BCFT$_2$ can be constructed by putting an AdS$_2$ end-of-the-world (EOW) brane in an AdS$_3$ spacetime. A holographic boundary RG flow is triggered by coupling a matter field to the EOW brane \cite{Takayanagi_2011_bcft, Fujita_Takayanagi_Tonni_2011}. Here one requires the matter field to satisfy the null energy condition (NEC), which implies that the matter is just ordinary matter with positive mass and energy. Since the gravity is an attractive force and repulsive effects are absent, matter being normal (satisfying NEC) immediately implies that the EOW brane can be only bended in one direction but not the other \cite{Omiya_2021}. This attractive nature of normal matter in gravity is the origin of the monotonically decreasing behavior of the holographic $g$-function, though it was not stated in this way in its original proof \cite{Fujita_Takayanagi_Tonni_2011}. The same principle has been applied to argue various kinds of monotonicity in the AdS/CFT correspondence. See, e.g. Ref.~\cite{Wei_2024_Crosscap, Fujiki_Kosei_Kanda_Kohara_Takayanagi_2025}. 

On the other hand, constructing non-unitary boundary conditions in holography sometimes require abnormal matter, i.e. those who do not satisfy NEC and hence have negative mass. One such example was presented in Ref.~\cite{2026_Takayanagi}. In Ref.~\cite{2026_Takayanagi}, the author use an imaginary-valued scalar field localized on the EOW brane to trigger an imaginary generalization of the holographic marginal RG flow to construct non-unitary boundary conditions. In this setup, the boundary condition $B$ and its conjugate $\bar{B}$ always come in pairs. The crucial point is that the imaginary-valued scalar field does not satisfy NEC and is hence abnormal. The force between such matter is repulsive, and hence can bend the brane in the other direction \cite{Omiya_2021}. In other words, it breaks the attractive nature of the gravity. 
With an RG flow triggered by this kind of matter, 
$|g|$ would rather increase than decrease. 

In this way, in this simple AdS/CFT toy model, implementing an RG flow compatible with the existence of non-unitary boundary conditions necessarily mixes up normal attractive matter and abnormal repulsive matter, and hence the monotonicity of the $g$-function is broken. 

This, however, does not imply that one cannot construct constrained RG flows under which certain quantities related to the $g$-function is monotonic. Exploring such constrained RG flows would be an interesting direction to explore. 

Before ending this section, we would like to comment on one concern of the above construction, where one uses abnormal repulsive matter to construct non-unitary boundary conditions in AdS/CFT \cite{2026_Takayanagi}. Such construction will lead to traversable wormhole on the AdS side, whose understanding from the CFT point of view remains puzzling.

\bibliography{ref}

\end{document}